\documentclass[11pt]{article}
\usepackage{style}
\usepackage{setspace}

\begin{document}

\title{Simulation and evaluation of local daily temperature and precipitation series derived by stochastic downscaling of ERA5 reanalysis}

\date{}

\author{Silius M. Vandeskog\(^1\) and Thordis L. Thorarinsdottir\(^1\) and Alex Lenkoski\(^1\)}

\maketitle

\(^1\) Norwegian Computing Center, P.O.Box 114 Blindern, Oslo, NO-0314, Norway

\begin{abstract}
Reanalysis products such as the ERA5 reanalysis are commonly used as proxies for observed atmospheric conditions. These products are convenient to use due to their global coverage, the large number of available atmospheric variables and the physical consistency between these variables, as well as their relatively high spatial and temporal resolutions. However, despite the continuous improvements in accuracy and increasing spatial and temporal resolutions of reanalysis products, they may not always capture local atmospheric conditions, especially for highly localised variables such as precipitation. This paper proposes a computationally efficient stochastic downscaling of ERA5 temperature and precipitation. The method combines information from ERA5 and surface observations from nearby stations in a non-linear regression framework that combines generalised additive models (GAMs) with regression splines and auto-regressive moving average (ARMA) models to produce realistic time series of local daily temperature and precipitation. Using a wide range of evaluation criteria that address different properties of the data, the proposed framework is shown to improve the representation of local temperature and precipitation compared to ERA5 at over 4000 locations in Europe over a period of more than 70 years.
\end{abstract}

\section{Introduction}

Climate impact studies commonly aim to answer questions about local impacts and thus require time series of local atmospheric variables, often over a period of several decades. When assessing historical and current impact, reanalysis products such as the ERA5 reanalysis produced by the European Centre for Medium-Range Weather Forecasts (ECMWF) offer a useful and convenient estimate of historical atmospheric conditions. Reanalysis products provide a global-scale physically consistent description of Earth's climate with information on a large number of variables, including variables that are challenging to observe directly. However, for highly localised observable variables, such as precipitation, several studies have shown that ERA5 is not always able to accurately capture local conditions \citep[e.g.,][]{Chen&2024, Lavers&2022, Liu&2024}. For example, \cite{Lavers&2022} perform a global evaluation of ERA5 precipitation against station observations using a nearest neighbour approach. They find that ERA5 generally has a wet bias, with the smallest errors occurring during winter in the Extratropics (e.g., in Europe) and the largest errors in the Tropics (e.g., across the Maritime Continent). 

Downscaling methods generally fall in two classes, dynamical and statistical downscaling, with statistical downscaling commonly used when the aim is to obtain descriptions of local conditions. \cite{MaraunEtAl2019Statisticaldownscalingskill} synthesize a comprehensive comparison of various statistical downscaling methods at 86 selected locations in Europe \citep{GutierrezEtAl2019experiment} that includes, among other things, specific studies on extremes \citep{HertigEtAl2019extremes} and temporal variability \citep{MaraunEtAl2019temporal}. The overall conclusion from these studies is that no single method will outperform all others for every conceivable situation. Weather generators simulate local weather sequences by modelling the statistical properties of the local weather. Thus, the short term (e.g., daily) variations in the simulations may not directly correspond to those in the coarser resolution data. However, such methods often perform well, especially if the general statistical properties of the local weather are well described by the training data, e.g., when target-resolution gridded data products are available for training \citep{YuanEtAl2019NewApproachBias, YuanEtAl2021Bridgingscalegap}. 

Perfect prognosis approaches establish a statistical relationship between coarser-scale climate descriptors and the local weather, preserving the short term variations in the coarser scale data, while model output statistics refers to methods that construct a direct statistical relationship between the coarser scale data and the local data. The performance of perfect prognosis approaches crucially depends on the model assumptions, including distributional assumptions, and the chosen climate descriptors.  Model output statistics is found to generally perform well if the predictors have a resolution close to the target resolution \citep{HertigEtAl2019extremes, MaraunEtAl2019Statisticaldownscalingskill}. The statistical relationship between the predictors and the response in perfect prognosis and model output statistics approaches is classically constructed using standard regression models. Recently, machine learning alternatives have been shown to perform well \citep[e.g.,][]{JiangEtAl2021, LiuEtAl2023, SachindraEtAl2018, SebbarEtAl2023, ZhuEtAl2023}. However, these methods usually only provide deterministic downscaling results, ignoring the additional random variability associated with the finer scale, and are compared against standard linear regression. Alternatively, modern regression approaches using regression splines that allow for non-linear relationships between the predictors and the response \citep{Wood2003ThinPlateRegression, Wood2017GeneralizedAdditiveModels} have been shown to perform well and, potentially, outperform neural network approaches \citep[e.g.,][]{NacarEtAl2022, SanEtAl2023}.

The aim of the current study is to construct a computationally efficient approach to obtain simulations of local daily temperature and precipitation series at any given location based on ERA5 and available local observation data. For this, we combine elements of model output statistics and weather generators. At locations where local data is available, we establish a statistical relationship between ERA5 temperature and precipitation and the local weather using generalized linear models with regression splines under appropriate distributional assumptions for each weather variable. Additionally, the models include a seasonal component as noted by \cite{MaraunEtAl2019Statisticaldownscalingskill}. We then combine these relationships with autoregressive simulation models in order to obtain simulations with realistic temporal correlation structure.

For locations without local observations, an alternative would be to use a regional model, estimated based on all available local observations in a region.
As in \cite{RobertsEtAl2019}, we find that elevation plays an important role in correcting biases in such models.
We thus use as predictors three different summary statistics describing the local and the corresponding ERA5 grid cell elevation, in addition to seasonality and latitude/longitude information.
Furthermore, we find that the regional approach may be substantially improved by augmenting the regional model with local estimates based on a specific set of donor locations in the vicinity of the location of interest.
Similar techniques are commonly used in hydrology to obtain hydrological simulations in ungauged catchments \citep[e.g.,][]{ArsenaultEtAl2023, GuoEtAl2021, KjeldsenEtAl2014}.
This yields a multi-model ensemble of local weather simulations, where the individual ensemble members vary due to both the estimated randomness at the local scale and in the specific local models used for the simulations.
Since simulating additional ensemble members comes at very low computational costs, the methods can easily generate ensembles of any size.

We apply and compare different variants of our method at over 4000 locations in Europe over a period of over 70 years.
Our method is also compared to a convection-permitting regional climate model (CPRCM) with kilometre-scale resolution, from the CORDEX-FPSCONV initiative, to evaluate its performance relative to one of the current state-of-the-art dynamical downscaling methods.
Similar studies commonly only assess the performance of downscaling methods using deterministic performance measures such as the root mean squared error (RMSE), the mean absolute error (MAE) or mean bias \citep[e.g.,][]{NacarEtAl2022, RobertsEtAl2019, SebbarEtAl2023, ZhaoEtAl2017, ZhuEtAl2023}, thus lacking an assessment of distributional properties.
For few study locations, it may be feasible to directly assess summary statistics such as mean, standard deviation and skewness at individual locations \citep{SanEtAl2023}.
Over many locations, this is no longer feasible and mean or relative bias of summary statistics may be evaluated.
Here, the focus may be on general summaries \citep{SachindraEtAl2018}, or summaries specifically related to temporal variability \citep{MaraunEtAl2019temporal}, extremes \citep{HertigEtAl2019extremes} or precipitation occurrence \citep{Liu&2024}.

A second contribution of this paper is a set of new evaluation criteria for simulated daily temperature and precipitation time series.
Specifically, we suggest that the simulations should correctly reflect the distributional properties of the weather variables at shorter and longer time scales.
Thus, we propose to evaluate distributions of summaries such as weekly and monthly mean and standard deviation, as well as daily and weekly differences, in addition to more standard evaluation of daily observations, tail properties and precipitation occurrence.
Distributions of observed summaries are compared against distributions of simulated summaries using the integrated quadratic divergence (IQD), also called the Cram\'er distance, which is the score divergence associated with the continuous ranked probability score proper scoring rule \citep{ThorarinsdottirEtAl2013UsingProperDivergence}.
It has previously been used for similar distributional evaluation in, e.g., \cite{YuanEtAl2019NewApproachBias}, \cite{ThorarinsdottirEtAl2020}, \cite{VeigaYuan2021}, \cite{YuanEtAl2021Bridgingscalegap}, \cite{DunnEtAl2022} and \cite{PengEtAl2022}.
Other evaluation criteria are, equivalently, based on proper scoring rules, ensuring meaningful conclusions from a decision-theoretic viewpoint \citep{GneitingEtAl2007StrictlyProperScoring}.
That is, all evaluation criteria are constructed such that, in expectation, a simulation model based on the true data distribution would have optimal performance. 

The remainder of the paper is organised as follows.
The data sets used in the analysis are introduced in Sect.~\ref{sec:data}, followed by a description of both the downscaling models and the evaluation criteria in Sect.~\ref{sec:model}. The results are presented in Sect.~\ref{sec:results}, with a discussion provided in Sect.~\ref{sec:discussion} and the conclusions summarised in Sect.~\ref{sec:conclusions}. 

\section{Data}%
\label{sec:data}

We downscale temperature and precipitation data over Europe from the ERA5 reanalysis data set~\citep{HersbachEtAl2020ERA5globalreanalysis} which is freely available from the Copernicus Climate Data Store~\citep{cds_era5}.
The data is downloaded at a spatial resolution of \(0.25^\circ \times 0.25^\circ\) and a temporal resolution of 1 hour.
The ERA5 reanalysis data are downscaled using local weather station data from the freely available Global Surface Summary of the Day (GSOD) data set~\citep{SparksEtAl2017GSODRGlobalSummary, gsod23}, which contains daily records of surface observations from more than 20,000 weather stations covering the globe, with the oldest stations going back to 1929.
We extract all available temperature and precipitation observations from the time period 1950-2024 over Europe.

\begin{figure}
  \centering
  \includegraphics[width=.95\linewidth]{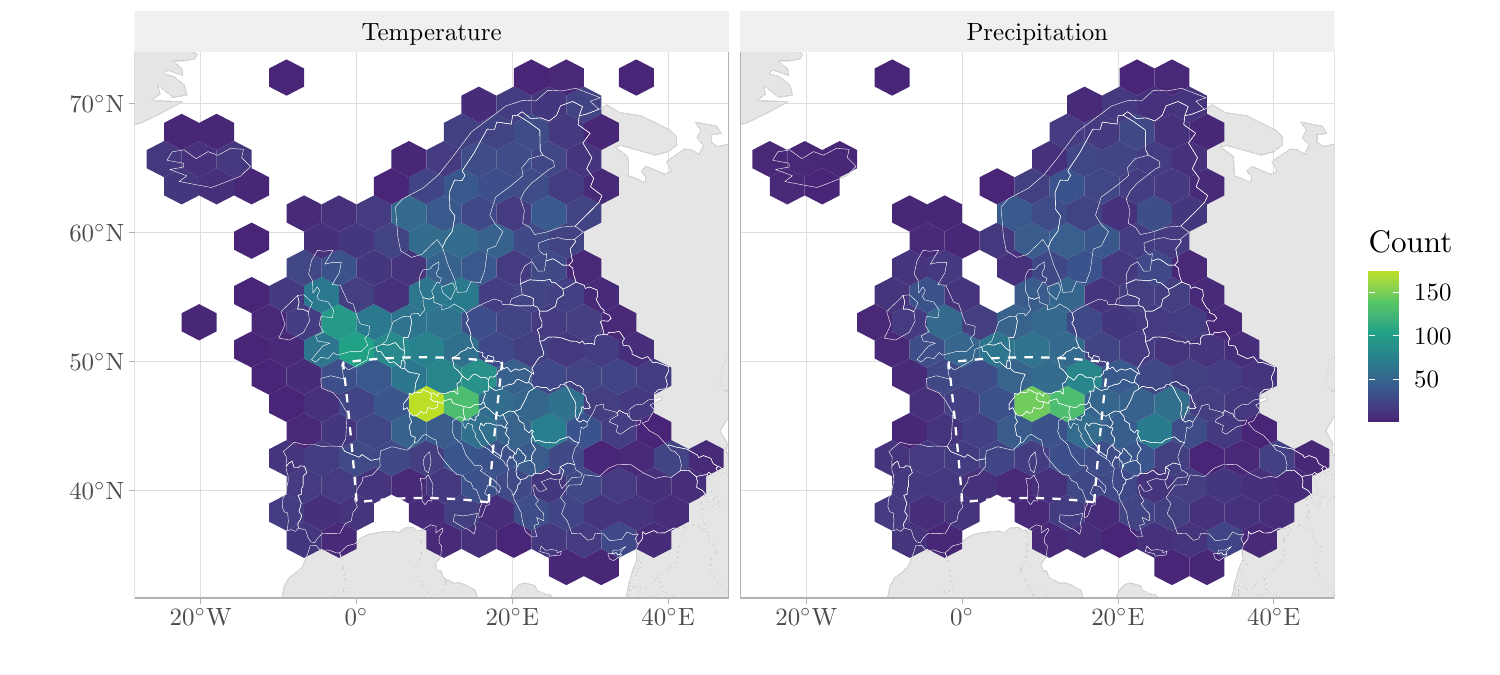}
  \caption{%
    Maps displaying the spatial density of weather stations, shown as counts on a hexagonal tiling of Europe.
    The left subplot displays the density of stations recording temperature, while the right subplot displays the density of stations recording precipitation.
    The boundary of the ALP-3 domain is displayed using a dashed line.
    }%
  \label{fig:map}
\end{figure}

Most of the daily precipitation data from GSOD come with a quality code describing the degree to which the observations are trustworthy.
We perform quality control of the data by excluding all daily precipitation observations with low quality, i.e., quality code H or I.
We then remove all precipitation data from weather stations with less than 200 precipitation observations, and all temperature data from weather stations with less than 200 temperature observations.
Finally, we remove all precipitation data from weather stations with less than 40 unique daily precipitation values.
This results in a data set consisting of \(4480\) unique weather stations, with all of these stations measuring daily mean temperature and \(3424\) stations measuring daily precipitation.
Figure~\ref{fig:map} displays the spatial distribution of these \(4480\) weather stations.
Since the station data have a daily temporal resolution, we aggregate the hourly ERA5 data in time to a daily resolution and develop our models at the daily scale.

To compare our downscaling method against the state-of-the-art within dynamical downscaling, we also download daily temperature and precipitation data from a CPRCM with kilometre-scale resolution.
There have been developed many different CPRCMs, covering different spatial domains.
We compare our downscaling method against a CPRCM over the ALP-3 domain~\citep{coppola18_europ_medit, ban21_i, pichelli21}.
This domain has a horizontal grid spacing of \(2.2\)--\(3\) km, and it covers a large part of central Europe, varying in altitude and climate, see Figure~\ref{fig:map}.
Using available data from the Earth System Grid Federation (ESGF), we then select the CPRCM version with greatest temporal coverage for both temperature and precipitation. 
The resulting CPRCM is based on the CNRM-AROME41t1 regional model~\citep{lucas-picher21_convec, lucas-picher23_evaluat_cnrm_arome4_north_europ}, driven by the ERA-Interim reanalysis data set~\citep{dee11_era_inter}, for the years 1982--2018.

We compare our stochastic downscaling method against the dynamical downscaling method at GSOD weather stations inside the ALP-3 domain.
To avoid boundary effects, we only compare the downscaling methods at GSOD weather stations that are more than one degree latitude away from the northernmost or southernmost point of the ALP-3 domain, and more than one degree longitude away from the westernmost or easternmost point of the ALP-3 domain.
This results in a total of \(931\) weather stations with temperature data.
\(797\) of these also contain precipitation data.

In addition to the weather data, we also rely on freely
available elevation data from the ETOPO Global Relief
Model~\citep{etopo22}, which contains global elevation data on a
\(15 \times 15\) arc-second spatial resolution, corresponding to a
spatial resolution of approximately 450~m.

Since our downscaling models are used for describing the dependence between spatially gridded data and point data, we use as covariates elevation data from the location to which we are downscaling, and the ERA5 grid cell from which we are downscaling.
Specifically, we use three elevation covariates given by the elevation at the location to which we are downscaling, the difference between the point elevation and the average elevation inside the corresponding ERA5 grid cell of interest, and the standard deviation of all elevation observations inside the corresponding ERA5 grid cell.
Of these three elevation covariates, we consider elevation difference to be the most important.
This is because temperature and precipitation are known to be strongly correlated with elevation.
Thus, when there is a large difference between the elevation of a weather station and the mean elevation of its corresponding ERA5 grid cell, we expect the ERA5 weather variables to be poor predictors for the weather station observations.

The elevations and elevation differences tend to be small. However,
for certain valleys and mountain tops they can reach values of several
thousand metres. The observed precipitation intensities are also
heavy-tailed, with a median of 2~mm/day and a maximum of close to
500~mm/day. For this reason, we standardise the precipitation
intensities and station elevations, using the log-transformation
\(x \rightarrow \log(x + 1)\), where we add \(1\) inside the logarithm
to ensure that precipitation intensities close to zero do not get
transformed to highly negative values. We do not standardise the
elevation differences, as they are less heavy-tailed than the station
elevations.
All covariates used in our downscaling models are listed in Table~\ref{tab:covariates}.

\section{Methods}%
\label{sec:model}

For any coordinate \(\bm s \in \mathbb R^2\) on Earth, we are
interested in a realisation of the local daily weather, either temperature
or precipitation, given as a time series \(\bm y(\bm s) = (y_1(\bm s),
y_2(\bm s), \ldots, y_T(\bm s))^\top\) of length \(T\). To obtain such
as realisation, we estimate the conditional distribution \(\bm y(\bm
s) \mid \bm X(\bm s)\), where \(\bm X(\bm s) = (\bm x_1(\bm s), \bm
x_2(\bm s), \ldots, \bm x_T(\bm s))\) is a multivariate time series of
\(M\) different covariates. To estimate this conditional distribution,
we rely on observations of the local weather, \(\mathcal Y = \{\bm
y(\bm s_i)\}_{i = 1}^N\), from a set of \(N\) different weather
stations, located at \(\bm s_1, \bm s_2, \ldots, \bm s_N\).

\subsection{Spatially varying models of daily temperature and precipitation series}

\begin{figure}[t]
  \centering
  \includegraphics[width=.95\linewidth]{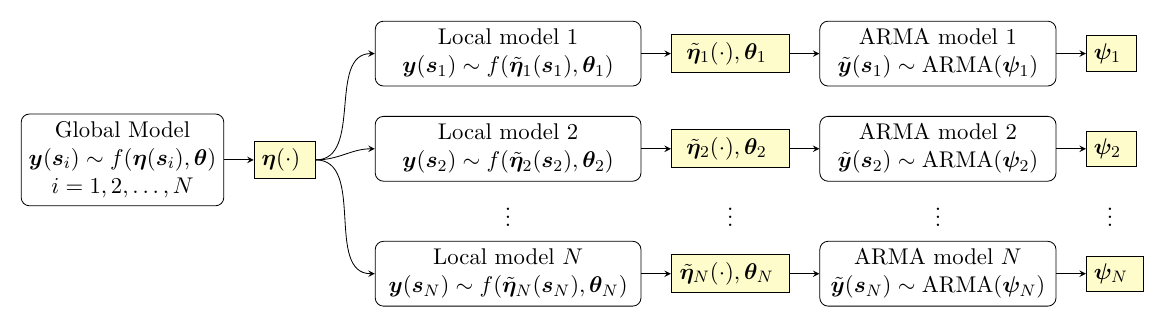}
  \caption{%
    A flowchart describing our proposed three-step downscaling model framework.
    First, the full data set is modelled jointly by a global model.
    Then, the non-linear predictor from the global model is used as a fixed offset in \(N\) different local models, one for each location.
    The local models at each location are then used to transform the data at that location to a Gaussian process, and a local ARMA model is used to model the temporal dependence of the Gaussian process.
  }%
  \label{fig:model_overview}
\end{figure}

We downscale daily precipitation amounts and daily mean 2~m
temperatures. While the downscaling is performed slightly differently for each
climate variable, the general framework is the same, and
is based on a three-step model that combines generalised additive
models (GAMs) and auto-regressive moving average (ARMA) models. The
model is visualised by the flowchart in
Figure~\ref{fig:model_overview}, and explained in detail below. Our
approach builds on common statistical downscaling
methods~\citep[e.g.,][]{MaraunEtAl2017StatisticalDownscalingBias,
  MaraunEtAl2019Statisticaldownscalingskill}.

In a first step of three, we fit a GAM model to all the local observations \(\mathcal Y\) using gridded weather data from ERA5 and orographic data as covariates, i.e.\
\begin{equation}
  \label{eq:global-gam}
  \left[y_t(\bm s_i) \mid \bm x_t(\bm s_i)\right] \sim f\left(\eta(\bm
    s_i), \bm \theta\right), \quad i = 1, 2, \ldots, N,\ t = 1, 2, \ldots, T,
\end{equation}
where \(f(\cdot)\) is the probability density function of some
reasonable probability distribution and \(\eta(\bm s_i)\) is a non-linear
predictor, built using information from the covariates (\(\bm x_t(\bm
s_i)\)), and \(\bm \theta\) contains a set of standard deviation parameters.
Specifically, for daily temperature, $f$ is the Gaussian density with mean  \(\eta(\bm s_i)\) using the identity link function and standard deviation \( \theta\).

The probability distribution of daily precipitation contains a point mass at zero, which makes downscaling more complicated.
Here, we separate the downscaling in two distinct parts, the probability of precipitation occurrence and the conditional precipitation intensity, given that precipitation occurs.
This separation into occurrences and conditional intensities is a common approach for modelling and simulating precipitation~\citep[e.g.,]{Richardson1981Stochasticsimulationdaily,  EvinEtAl2018Stochasticgenerationmulti, WilcoxEtAl2021StochastormStochasticRainfall}.
Specifically, precipitation occurrence is assumed to follow a Bernoulli distribution with parameter \(\eta(\bm s_i)\), using a logit link function.
The precipitation amount on a wet day is assumed to follow a gamma distribution with mean \(\eta(\bm s_i)\), using a logarithmic link function, and scale parameter \( \theta \), so that the variance equals \(\eta(\bm s_i) \theta\). 

When performing downscaling over a large spatial domain, it might be unreasonable to assume that the dependence between \(\bm y(\bm s)\) and \(\bm X(\bm s)\) is constant in space.
Therefore, in the second step of the three-step model, we model residual variability at each of the \(N\) locations where we have local data, using a set of local GAMs.
While the local models are similar to the global model~\eqref{eq:global-gam}, they rely on using the non-linear predictor from the global model as a fixed offset, i.e.\
\begin{equation}
  \label{eq:local-gam}
  \left[y_t(\bm s_i) \mid \bm x_t(\bm s_i)\right] \sim f\left(\tilde \eta_i(\bm s_i),
    \bm \theta_i\right), \quad t = 1, 2, \ldots, T,
\end{equation}
where \(\tilde \eta_i(\bm s) = \eta(\bm s) + \eta_i(\bm s)\) is the sum of the fixed offset \(\eta(\cdot)\) from the global model and a local non-linear predictor \(\eta_i(\cdot)\) that must be estimated. 

Finally, in the third and last step, we add a temporal dependence structure into our model, by standardising the residuals of the local models and fitting different ARMA models to them.
The standardisation is achieved using the probability integral transform (PIT), which allows us to transform any continuous random variable \(y\) into a standardised Gaussian distributed random variable \(\tilde y\), using the transform \( \tilde y = \Phi^{-1}(F(y)), \) where \(F\) is the cumulative distribution function (CDF) of \(y\) and \(\Phi^{-1}\) is the inverse CDF of the standardised Gaussian distribution.
Since we estimate the conditional marginal distribution of \(y_t(\bm s_i)\) with the local model in~\eqref{eq:local-gam}, we are able to transform \(\left[y_t(\bm s_i) \mid \bm x_t(\bm s_i)\right]\) into the approximately standardised Gaussian random variable \(\tilde y_t(\bm s_i)\).

It is not possible to transform the binary precipitation occurrences into Gaussian random variables in this way.
Therefore, ARMA models are only fitted to the residuals of the local temperature and precipitation intensity models and we model the temporal dependence structure of the precipitation occurrences using a first-order Markov chain.
This is a common approach for modelling and simulating daily precipitation occurrences and intensities~\citep[e.g.,][]{Richardson1981Stochasticsimulationdaily, MaraunEtAl2017StatisticalDownscalingBias}.
In essence, instead of using a single GAM to model the occurrence probability of precipitation on day \(t\), we use two separate GAMs.
One GAM models the occurrence probability at day \(t\) given that day \(t-1\) was wet, while a second GAM models the occurrence probability given that day \(t-1\) was dry.
Both models condition on the ERA5 precipitation values on both day \(t-1\) and \(t\). 

Combining all three modelling steps into one model yields a conditional probability density function which we denote
\(
  f(\bm y(\bm s_i); \tilde{\bm \eta}_i(\bm s_i), \bm \theta_i, \bm \Psi_i),
\)
where \(\bm \Psi_i\) contains all parameters of the ARMA model from step 3, \(\bm \theta_i\) contains all additional GAM parameters from the local GAM in step 2 and \(\tilde{\bm \eta}_i\) contains the non-linear predictors from the two GAMs in step 1 and 2.

\begin{table}[t]
  \centering
  \caption{%
    Covariates used for the downscaling models. The ``Models'' column describes in which downscaling model the different covariates are used.
    ``GP'' and ``GT'' indicate that the covariate is present in all global precipitation and temperature GAMs, respectively.
    Similarly, ``LP'' and ``LT'' indicate that the covariate is present in all local precipitation and temperature GAMs, respectively.
    The ``Description'' column provides more details about each covariate, and about how their effects are modelled.%
  }%
  \label{tab:covariates}
  \renewcommand*{\arraystretch}{1.5}
  \begin{tabular}{p{.12\linewidth}lp{.65\linewidth}}
    \toprule
    Covariate & Models & Description  \\
    \midrule
    Temperature & GP, GT, LP, LT & \setstretch{.5} Thin plate spline~\citep{Wood2003ThinPlateRegression} on ERA5 daily mean temperature at
        2~m elevation, with an optional additional constant offset term used for downscaling temperature. \\
    Precipitation & GP, GT, LP, LT & Thin plate spline on ERA5 daily aggregated precipitation, transformed with the function \(x
        \rightarrow \log(x + 1)\). \\
    \setstretch{0.7} Precipitation \qquad \qquad occurrence & GP, LP & \setstretch{.5} Linear effect on the Boolean variable that describes if the ERA5 daily precipitation amount is nonzero or not. This covariate is only used in the global and local precipitation occurrence models. \\
    Day & GP, GT, LP, LT & Cyclic cubic regression spline~\citep{Wood2017GeneralizedAdditiveModels} on the day of the year (1-366). \\
    Elevation & GP, GT & Thin plate spline on station elevation. \\
    \setstretch{0.7} Elevation \qquad difference & GP, GT & \setstretch{.5} Thin plate spline on the difference between station elevation and the mean elevation inside the ERA5 grid cell. \\
    Elevation SD & GP, GT & \setstretch{.5} Thin plate spline on the standard deviation of all elevation observations within the ERA5 grid cell. \\
    \setstretch{0.7} Longitude/\qquad Latitude & GP, GT & \setstretch{.5} Two-dimensional spline that lives on a sphere, taking in longitude and latitude values~\citep{Wahba1981SplineInterpolationSmoothing,Wood2003ThinPlateRegression}. \\
    \bottomrule
  \end{tabular}
\end{table}

An overview of the covariates used for each modelling component is given in Table~\ref{tab:covariates}, together with information about the GAMs in which they are used, and how their effects are modelled.
All models include ERA5 information on both temperature and precipitation and a seasonal covariate describing the day of the year.
The global models in step one additionally include information on local elevation from the ETOPO Global Relief model, the elevation difference between the current location and the corresponding ERA5 elevation, the standard deviation (SD) of elevation observations inside the ERA5 grid cell, as well as the latitude and longitude of the current location.
We assume that most covariates have a nonlinear effect on the weather variable to be downscaled.
For this reason, effects from all continuous covariates are modelled using various spline functions \citep{Wood2003ThinPlateRegression, Wood2017GeneralizedAdditiveModels}.
The temperature covariate is treated differently when downscaling temperature and precipitation.
Specifically, we assume that the downscaled temperature should be similar to the gridded temperature, and we therefore model the effect of the gridded temperature using a constant offset plus a spline function.
In other words, the effect of gridded temperature inside the linear predictors for temperature are given by the transformation \(x \rightarrow x + s(x)\), where \(s(x)\) is a spline function.
A similar offset is not used for downscaling the precipitation intensities.
The local Markov chain occurrence models use ERA5 data from both day \(t - 1\) and day \(t\) for describing the occurrence probabilities for day \(t\).
All other models only use ERA5 data from day \(t\) for modelling the local weather at day \(t\).

All GAMs are implemented using the \texttt{bam()} function of the \texttt{mgcv} package~\citep{Wood2017GeneralizedAdditiveModels} in \texttt{R}, while all ARMA models are fitted to data using the \texttt{auto.arima()} function from the \texttt{forecast} package~\citep{HyndmanEtAl2008AutomaticTimeSeries} in \texttt{R}.
This function automatically estimates the order of the AR- and MA-components within the ARMA models, which can differ for each of the local models.

\subsection{Downscaling to out-of-sample locations}

While the global model~\eqref{eq:global-gam} makes it possible to estimate the conditional distribution of \(\bm y(\bm s)\) for any location \(\bm s\), our three-step model only allows for downscaling to one of the \(N\) locations where we already have observations of the local weather.
To simulate local weather at a new location \(\bm s_0\), we therefore build upon ideas from regionalisation methods in hydrology \citep[e.g.,][]{ArsenaultEtAl2023}.
That is, we approximate the conditional distribution of \(\bm y(\bm s_0)\) with an ensemble of draws from available models at a set of donor locations where we have observed data and can estimate the local models.
Specifically, we use the locations of the \(K\) weather stations that are closest to \(\bm s_0\) in Euclidean distance as donor locations.
To model \(\left[\bm y(\bm s_0) \mid \bm X(\bm s_0)\right]\), we then insert the covariates from \(\bm s_0\) into all the donor models to create the predictive ensemble \(\left\{\bm y^{(1)}(\bm s_0), \bm y^{(2)}(\bm s_0), \ldots, \bm  y^{(B)}(\bm s_0)\right\}\), consisting of \(B\) ensemble members, where the first \(B/K\) ensemble members are sampled using the local model at the first donor location, the next \(B/K\) ensemble members are sampled using the local model at the second donor location, and so on.

To simulate ensembles of temperature and precipitation data from one
specific donor model, we first simulate ensembles of time series from
the locally fitted ARMA models. We then transform these to have more
appropriate marginal distributions, using the global and local GAMs,
with covariates from \(\bm s_0\). Precipitation occurrences are
simulated separately, from their respective global and local GAMs.

In practice, when simulating downscaled temperature and precipitation, we often find that one or two
of the donor models behave considerably different than the rest. These might, e.g., produce ensemble
members with considerably different precipitation sums, temperature averages or temporal
autocorrelations. In a final post-processing step, we therefore remove all simulated
ensemble members from these donor models. To decide which donors to remove, we implement a simple
outlier detection method, inspired by how outliers are defined in box-plots.  First, we choose a
statistic, such as the overall standard deviation of a simulated ensemble member.  Then, for each
donor model, we compute the average of this statistic for all ensemble members from that donor. A
donor model is then considered as an outlier if its average statistic is outside the range
\([Q_{25} - 1.5 \times IQR, Q_{75} + 1.5 \times IQR]\), where \(Q_{25}\) and \(Q_{75}\) are the empirical first
and third quartiles of the \(K\) different average statistics, respectively, and \(IQR\) is the
empirical inter-quartile range \(IQR = Q_{75} - Q_{25}\). For precipitation, we detect outliers
using the overall cumulative precipitation sum as our statistic. For temperature, we detect outliers
using both the overall temperature average and the overall
temperature standard deviation. Having removed all ensemble members from a set of donor models, we then simulate more ensemble members from the remaining donor models, to ensure that our final ensemble has a total of \(B\) members.

\subsection{Evaluation}%
\label{sec:evaluation}

We downscale daily temperature and precipitation by simulating a multi-model ensemble of \(B = 150\) conditional time series for each weather variable, see Sect.~\ref{sec:results} below for details on how the ensemble is constructed.
For a thorough comparison of simulated and observed weather, we compute multiple different evaluation criteria each summarized by the average score over the \(N\) stations, see Table~\ref{tab:evaluation} for an overview.
Each criteria evaluates a different property of the simulations.
For example, the RMSE provides information on the overall performance.
However, it will not assess the temporal autocorrelation, precipitation occurrence, or if the ensemble spread is reasonable.
By focusing on multiple evaluation criteria, we achieve a better understanding of the overall properties of the different downscaling methods.
Most of our evaluation criteria are inspired by \citet{Heinrich_MertschingEtAl2024Validationpointprocess}, who compare observed and simulated data by computing different summary statistics of the data and then using proper scoring rules \citep{GneitingEtAl2007StrictlyProperScoring} to compare the marginal distributions of those summary statistics.

\begin{table}[t]
  \centering
  \renewcommand*{\arraystretch}{1.5}
  \caption{All evaluation criteria used for evaluating different properties of our downscaling models.}%
  \label{tab:evaluation}
  \begin{tabular}{lp{.87\linewidth}}
    \toprule
    Name & Description  \\
    \midrule
    RMSE & RMSE between the daily observations and the ensemble mean of the
           simulated data. \\
    MAE & Mean absolute error (MAE) between the daily observations and the ensemble median of the
          simulated data. \\
    ZPX & The squared error between the observed and the simulated relative frequency of zero precipitation, where we
          define every precipitation value below X~mm/day as a
          zero. We use ZP01 and ZP1 in this manuscript, which
          corresponds to thresholds of \(0.1\)~mm/day and \(1\)~mm/day
          respectively. \\
    QX & Quantile score that evaluates the X-percentile of the simulated
         data. We use Q95/Q99 for precipitation and Q01/Q99 for temperature, corresponding to
         the 1st, 95th and 99th percentiles. \\
    IQD & IQD between the marginal distribution of the observed data and the marginal distribution
          of the simulated ensemble. For precipitation data, we only compare the marginal
          distributions of non-zero precipitation values. \\
    WM & IQD between the marginal distributions of the weekly means of observed and 
         simulated data. \\
    WS & IQD between the marginal distributions of the weekly standard deviations of observed
         and simulated data. \\
    MM & IQD between the marginal distributions of the monthly means of observed
         and simulated data. \\
    MS & IQD between the marginal distribution of the monthly standard deviations of observed
         and simulated data. \\
    DX & IQD between the marginal distribution of all X-day differences for observed
         and simulated data. We use D1, D3 and D7 in this manuscript. \\
    \bottomrule
  \end{tabular}
\end{table}

We evaluate the ensemble mean against the observation using RMSE and the ensemble median against the observation using MAE, thus selecting the optimal ensemble summary for each criteria \citep{Gneiting2011MakingEvaluatingPoint}.
We evaluate the overall level of dry days by calculating the squared difference \((\hat{p}_y - \hat{p}_z)^2\) between the observed relative frequency of dry days \(\hat{p}_y\) and the simulated relative frequency of dry days \(\hat{p}_z\), which is the score divergence associated with the Brier score \citep{Brier1950, GneitingEtAl2007StrictlyProperScoring}.
We evaluate tail properties using the quantile score~\citep[e.g.,][]{Gneiting2011MakingEvaluatingPoint},
\begin{equation}
  S_{\tau}(F, y) = \left(I(y < F^{-1}(\tau)) - \tau\right) \times \left(F^{-1}(\tau) - y\right),
\end{equation}
where \(y\) is a weather station observation, \(I(\cdot)\) is an indicator function and \(F^{-1}(\tau)\) is a quantile of the predictive distribution \(F\) at probability level \(\tau \in (0,1)\).
This score can directly evaluate, e.g., heavy rainfalls or high/low temperatures.
Since the dynamical downscaling data and ERA5 reanalysis observations are deterministic, in the sense that only one ensemble member is available, we cannot create different predictive distributions for every day, which we can for the different downscaling models.
Therefore, to compare quantile scores between ERA5 or the CPRCM, and our downscaling models, we compute \(F^{-1}(\tau)\) as the empirical \(\tau\)-quantile of the entire time-series of ERA5/CPRCM observations, or the set of all simulated time-series from all ensemble members of a given downscaling model.

To evaluate the temporal autocorrelation, we compute differences between weather at day \(d\) and day \(d + n\), for different values of \(n\). Then, we compare the marginal distributions of these \(n\)-day differences between the observed data and the simulated data, using the integrated
quadratic distance (IQD)~\citep[e.g.][]{ThorarinsdottirEtAl2013UsingProperDivergence}
\begin{equation}
  \text{IQD}(\bm T_z, \bm T_y) = \int_{-\infty}^\infty \left(\hat F(t) - \hat G(t)\right)^2 \text{d} t,
\end{equation}
where \(\hat G\) is the empirical CDF of the computed summary statistics from the observed data,
\(\bm T_y\), while \(\hat F\) is the empirical CDF of the computed summary statistics from the
simulated data, \(\bm T_z\). The empirical CDF \(\hat F\), for the simulated data, is created by
pooling together the summary statistics from all \(B\) ensemble members. Similarly, we use the IQD to evaluate weekly and monthly means and standard deviations, as well as the overall distribution of the weather at a given location. 

As mentioned above, we summarise the skill of each model by computing average scores over all \(N\) locations. However, since different criteria produce values on different scales, it can be difficult to compare models simply by examining pairwise differences of the average scores. It is therefore common to standardise average scores by transforming pairs of average scores into so-called skill scores
\citep[e.g.][]{GneitingEtAl2007StrictlyProperScoring}
\begin{equation}
  \label{eq:skill-score}
  S_{\text{skill}}(S_1, S_0) = \frac{S_0 - S_1}{S_0} = 1 - \frac{S_1}{S_0}.
\end{equation}
In this setting, the model corresponding to \(S_0\) is typically considered as the base model, while the model corresponding to \(S_1\) is considered as the competitor.
The skill score is thus the relative difference in the scores, with respect to the base model, so that a skill score of 0.05 indicates that \(S_1\) is 5\% better than the base model.

\section{Results}%
\label{sec:results}

We apply the three-step model framework from Sect.~\ref{sec:model} for downscaling daily precipitation and daily mean 2~m temperature from ERA5 to any location within Europe over the time period 1950--2024.
Specifically, we downscale daily temperature and precipitation for each of the \(N\) available weather stations by simulating a multi-model ensemble of \(B = 150\) conditional time series for each weather variable.
For this, we use local models from the \(K\) nearest weather stations, with \(K \in \{5, 10, 15, 20, 25, 30\}\).
We then evaluate the predictive performance of our downscaling method by comparing the simulated ensemble with observed data from each station.
Our evaluation procedure is similar to leave-one-out cross-validation in that the locally fitted model for location \(i\) is not used for simulating local weather at that location.
The differences between true leave-one-out cross-validation and our analysis are further discussed in Sect.~\ref{sec:discussion}.
To test if all three modelling steps from Sect.~\ref{sec:model} provide value, we compare our three-step downscaling model with simpler models, where we stop after the first or the second of the three steps in Sect.~\ref{sec:model}.
We denote the models stopping after the first, the second and the third step as the global, the local and the full model, respectively.
To test the overall gains of downscaling, we also compare against the raw ERA5 data, and we compare against the chosen CPRCM within the ALP-3 domain.

Properties of the global GAMS used in our downscaling models are evaluated in Sect.~\ref{sec:global_models}.
Comparisons between ERA5 and different sub-models of our downscaling models, over all 4480 available weather stations, are presented in Sect.~\ref{sec:skill_scores},~\ref{sec:skill_score_trends} and~\ref{sec:time_series_plots}.
Finally, our stochastic downscaling models are compared against the chosen CPRCM in Sect.~\ref{sec:cprcm}.

The developed downscaling models are built to downscale weather to one specific location at a time, and not to capture the spatial correlation between multiple locations.
It might still be of interest to evaluate the spatial coherence of locally downscaled weather at multiple different locations.
We therefore evaluate spatial properties of the downscaling models in Appendix~\ref{app:spatial-coherence}.

\subsection{Global model fits}%
\label{sec:global_models}

\begin{figure}[t]
  \centering
  \includegraphics[width=.95\linewidth]{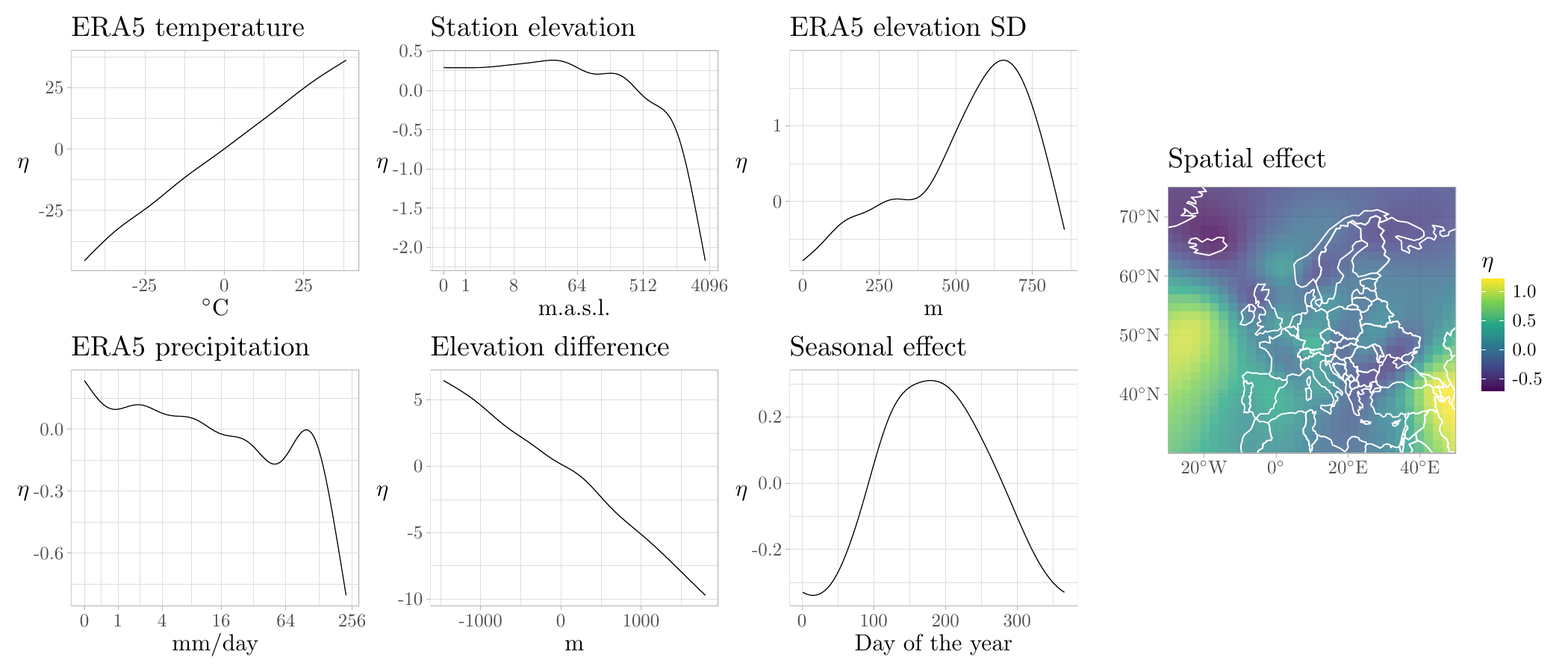}
  \caption{
    All fitted splines in the global temperature model.
  }%
  \label{fig:global_temp_model}
\end{figure}

\begin{figure}[t]
  \centering
  \includegraphics[width=.98\linewidth]{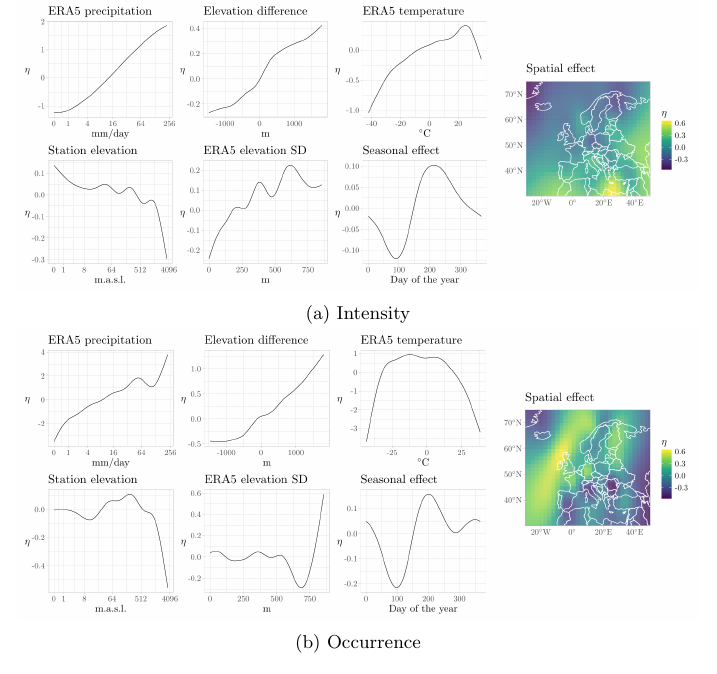}
  \caption{
    All fitted splines in the global precipitation intensity model (upper rows) and occurrence model (lower rows).
  }%
  \label{fig:global_precip_model}
\end{figure}

First, we investigate the estimated model parameters for the global models.
Figures~\ref{fig:global_temp_model} and~\ref{fig:global_precip_model} display all fitted splines in the non-linear predictors of the global GAMs for temperature and precipitation, respectively.
As expected, ERA5 temperature and the elevation difference between weather stations and ERA5 grid cells are the two most important covariates for downscaling temperature from ERA5.
The seasonal effect and the ERA5 precipitation effect are less important.
As seen in Figure~\ref{fig:global_precip_model}, positive elevation differences, i.e., weather stations that are higher than the mean altitude of their corresponding ERA5 grid cells, are also associated with both larger probabilities of precipitation occurring, and larger expected precipitation intensities.
Interestingly, ERA5 temperature is one of the covariates with the strongest effects in the global occurrence model.
This is likely due to extreme low and high temperatures tending to occur during dry conditions.
The global precipitation occurrence model also includes a covariate describing whether the daily ERA5 precipitation amount is positive or not.
This covariate increases the non-linear predictor by \(1.02\) when the daily ERA5 precipitation amount is positive.

Overall, Figures~\ref{fig:global_temp_model} and \ref{fig:global_precip_model} show that the statistical relationships between the predictors and the response are clearly non-linear, except for the relationships between the temperature response and ERA5 temperature and elevation difference, respectively, cf.\ Figure~\ref{fig:global_temp_model}.
This supports the need for a more flexible modelling framework than standard linear regression, and we see that the regression splines are able to capture a wide range of relationships.
Furthermore, the results support previous findings that elevation information \citep{RobertsEtAl2019} and seasonality corrections \citep{MaraunEtAl2019Statisticaldownscalingskill} are important. 

\subsection{Overall skill scores}%
\label{sec:skill_scores}

\begin{figure}[t]
  \centering
  \includegraphics[width=.95\linewidth]{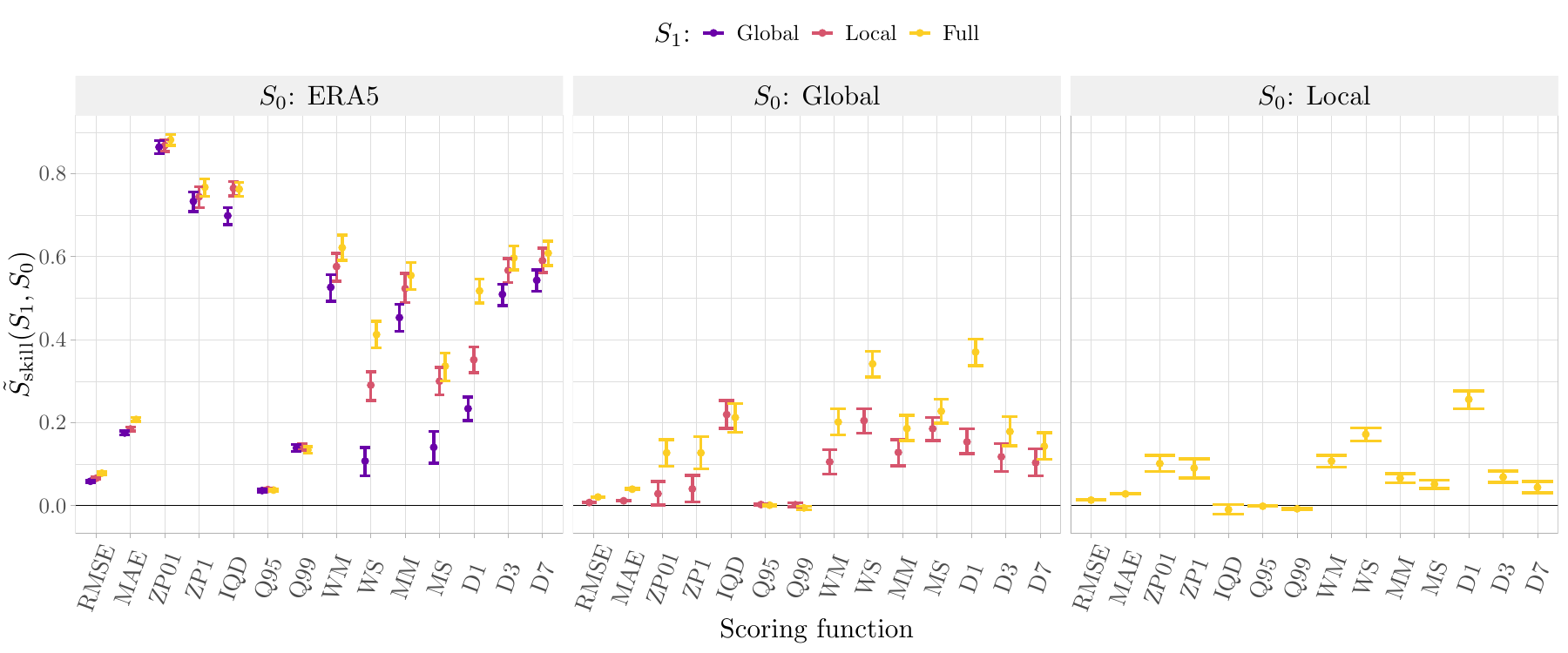}
  \caption{Skill scores comparing the four precipitation downscaling models, with bootstrapped
    \(95\%\) confidence intervals, for all evaluation criteria from
    Table~\ref{tab:evaluation}. Competitor models (\(S_1\)), are distinguished by different colours,
    while base models (\(S_0\)), are distinguished by the different subplots.}%
  \label{fig:precip_scores}
\end{figure}

\begin{figure}[t]
  \centering
  \includegraphics[width=.95\linewidth]{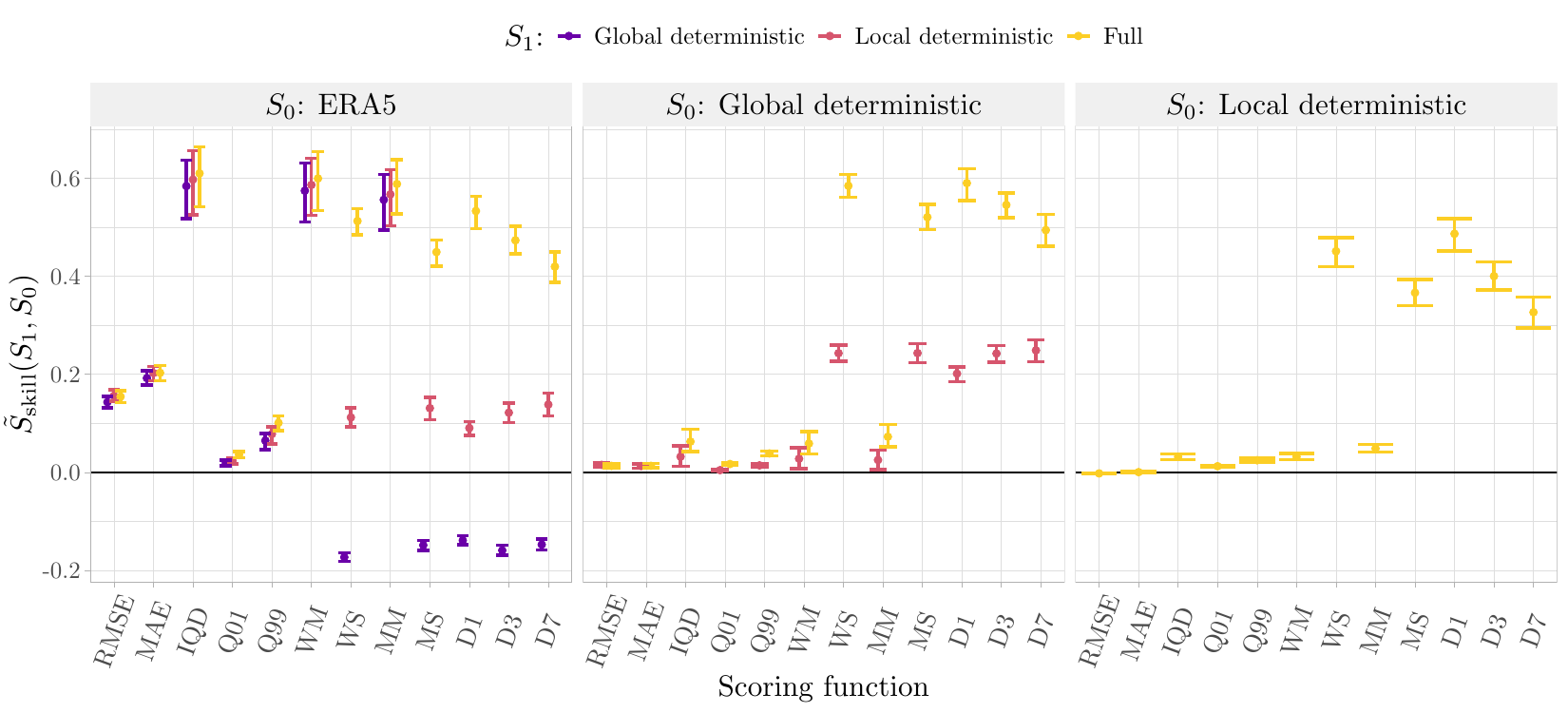}
  \caption{Skill scores comparing the four temperature downscaling models, with bootstrapped
    \(95\%\) confidence intervals, for all evaluation criteria from
    Table~\ref{tab:evaluation}. Competitor models (\(S_1\)), are distinguished by different colours,
    while base models (\(S_0\)), are distinguished by the different subplots.}%
  \label{fig:temp_scores}
\end{figure}

We use the skill score in~\eqref{eq:skill-score}, with all evaluation criteria in Table~\ref{tab:evaluation}, to choose the optimal number of donors, \(K\), for the local and the full downscaling models.
The results are displayed in Figures~\ref{fig:precip_K_scores} and~\ref{fig:temp_K_scores} in Appendix~\ref{app:evaluation}.
Based on these figures, we choose \(K = 15\) for precipitation and \(K = 10\) for temperature. We then compute skill scores for our different downscaling models, and we create bootstrapped \(95\%\) confidence intervals by resampling scores from the \(N\) different weather station locations with replacement.
The results are displayed in Figures~\ref{fig:precip_scores} and~\ref{fig:temp_scores}, for precipitation and temperature, respectively.
In Figure~\ref{fig:temp_scores}, we have exchanged the global and local models for two models denoted the local deterministic and global deterministic models.
This is because the time-independent Gaussian white noise that gets added when we simulate downscaled temperature ensembles substantially reduces the predictive performance (results not shown).
We therefore replace the two models with deterministic downscaling models that only rely
on the non-linear predictors of the global and local GAM models.

The results show that downscaling improves the skill compared to raw ERA5 data, for both temperature and precipitation, and for all chosen evaluation criteria.
The only exception is the global deterministic temperature downscaling method, which results in worse skill for all scores that focus on temporal autocorrelation.
The improvement over ERA5 is most considerable for precipitation occurrence.
The local and full models tend to outperform the global model, which implies that the use of local information from neighbouring weather stations can add additional skill when downscaling.
The full temperature downscaling model considerably outperforms its local deterministic variant, which implies that it is crucial to include temporal autocorrelation when downscaling temperature.
However, this effect is not visible in the deterministic scores RMSE and MAE, stressing the need for evaluation that specifically focuses on temporal autocorrelation. 

The full precipitation downscaling model outperforms the local model for most evaluation criteria.
Further examination shows that the increase in skill mostly stems from the added Markov chain structure of the occurrence models.
The temporal trends of the precipitation intensity models are weak at many locations, with almost half of the fitted intensity ARMA-models being identical to Gaussian white noise.

Our outlier removal method removes between 0 and 2 outlier donors for most locations.
This removal leads to a considerable improvement for most of the chosen evaluation criteria (results not shown).

\subsection{Skill score trends}%
\label{sec:skill_score_trends}

\begin{figure}[t]
  \centering
  \includegraphics[width=.95\linewidth]{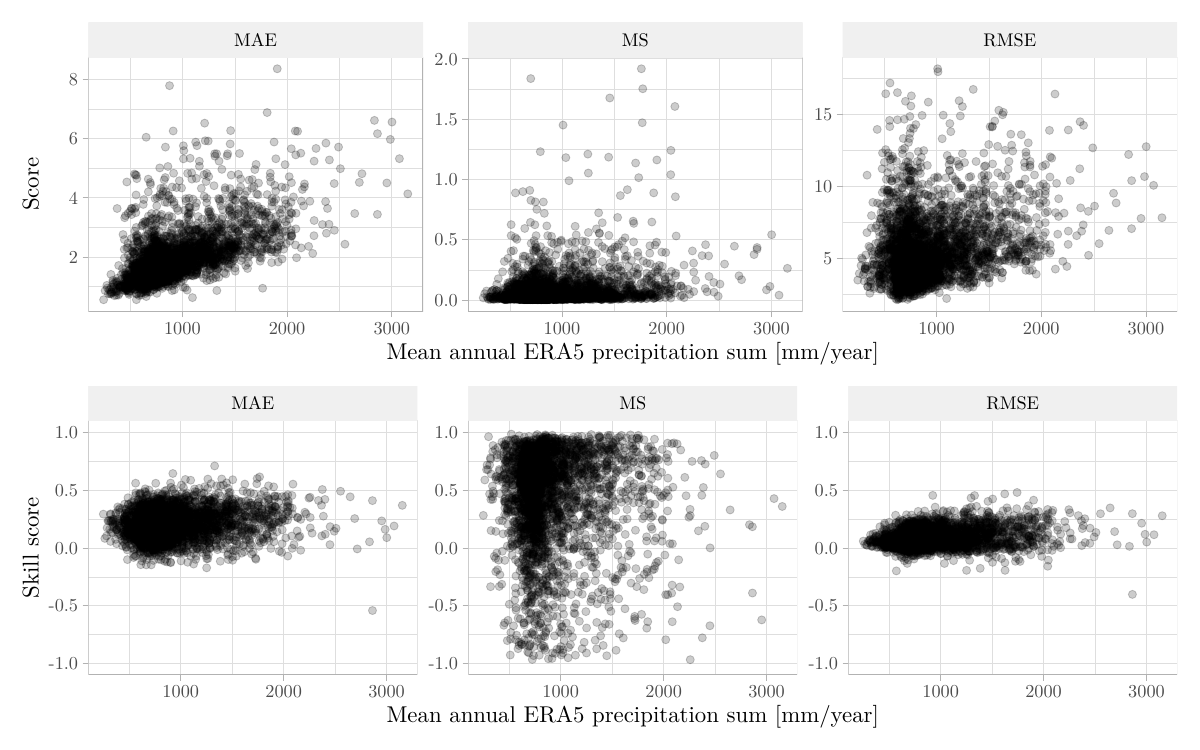}
  \caption{%
Scatter plots displaying scores (top row) and skill scores compared to ERA5 (bottom row) for the full precipitation downscaling model at all available weather stations plotted against the total mean annual ERA5 precipitation.
    For visualisation purposes, the \(0.02\%\) largest scores, and all skill scores below \(-1\), have been removed from each scatter plot.
  }%
  \label{fig:precip_scatter}
\end{figure}

Figure~\ref{fig:precip_scatter} displays scatter plots of scores and skill score, created using three of the evaluation criteria in Table~\ref{tab:evaluation}, for the full precipitation downscaling model.
The MAE and RMSE scores in the upper row display a strong correlation with the average annual precipitation sums at each station.
This highlights the importance of not relying on only one or two evaluation criteria.
RMSE and MAE describe absolute errors instead of relative errors, and absolute errors tend to be larger in areas with more precipitation.
Thus, overall model performance, with respect to RMSE or MAE, will be affected more by locations with large amounts of precipitation than by those with little precipitation.
This is not the case for the MS scores, which do not appear to be strongly correlated with the mean annual precipitation sums.
The skill score scatter plots in the lower row tell a different story.
Here, there are no correlations between the mean annual precipitation sums and the MAE/RMSE skill scores, whereas the MS skill scores appear to increase somewhat as the annual precipitation sum increases.
However, this apparent correlation does not seem to be very strong, and it could appear to be stronger than it actually is, due to the upper censoring at a skill score of 1.
One should therefore be careful about drawing too strong conclusions about the downscaling model from this. 
Figure~\ref{fig:precip_scatter} highlights the importance of examining both raw scores and skill scores, as the two might provide different types of information.
In this case, we find that, even though MAE and RMSE are strongly connected to the mean annual precipitation, this is the case for both the full downscaling model and ERA5, and the overall performance increase of the full model does not seem to be connected to the annual precipitation amounts.
The removal of large scores and small skill scores from some of the scatter plots does not alter the overall findings.

\begin{figure}[t]
  \centering
  \includegraphics[width=.95\linewidth]{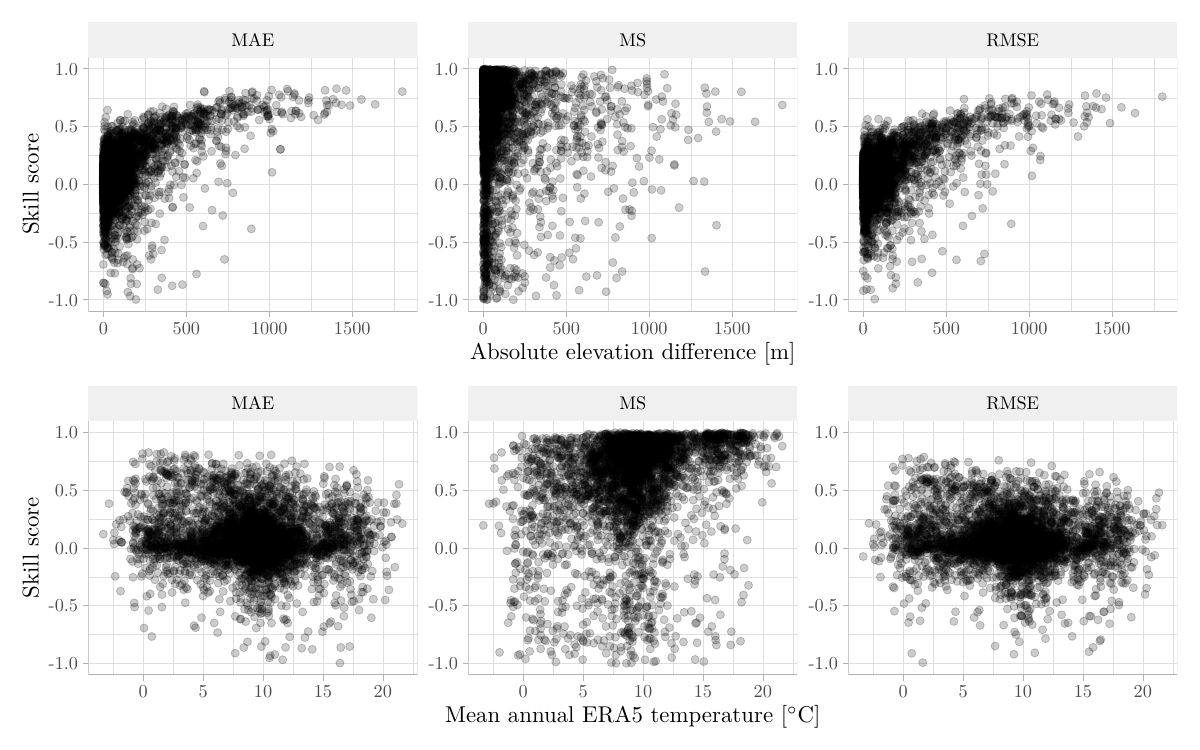}
  \caption{%
    Scatter plots displaying skill scores for the full temperature downscaling model compared to ERA5 at all available weather stations.
    The x-axes display the absolute elevation difference between weather stations and ERA5 grid cells (top row) or the mean annual ERA5 temperature (bottom row). 
    For visualisation purposes, all skill scores below \(-1\), have been removed from each scatter plot.
  }%
  \label{fig:temp_scatter}
\end{figure}

Figure~\ref{fig:temp_scatter} displays skill score scatter plots for the full temperature downscaling model, created using the same evaluation criteria as in Figure~\ref{fig:precip_scatter}.
There is considerable correlation between the elevation differences and MAE/RMSE skill scores.
This makes sense, as temperature is strongly correlated to elevation, meaning that ERA5 is expected to be biased as a predictor for local temperature in grid cells with large elevation differences.
However, there are no clear correlations between the elevation differences and MS skill scores.
The lower row of Figure~\ref{fig:temp_scatter} show that the MS skill scores appear to be weakly positively correlated with the mean annual ERA5 temperatures, while the MAE/RMSE skill scores appear to be weakly negatively correlated with the mean annual temperatures.
This implies that the full downscaling model is slightly better at improving temporal correlations, and slightly worse at removing biases, in warmer regions.
Once more, this highlights the importance of relying on multiple different criteria for model evaluation.
The removal of small skill scores from some of the scatter plots does not alter the overall findings.
Some of the patterns in Figures~\ref{fig:precip_scatter} and~\ref{fig:temp_scatter} are examined in more detail in Appendix~\ref{app:evaluation}.

\subsection{Visualising specific time series}%
\label{sec:time_series_plots}

\begin{figure}[t]
  \centering
  \includegraphics[width=.95\linewidth]{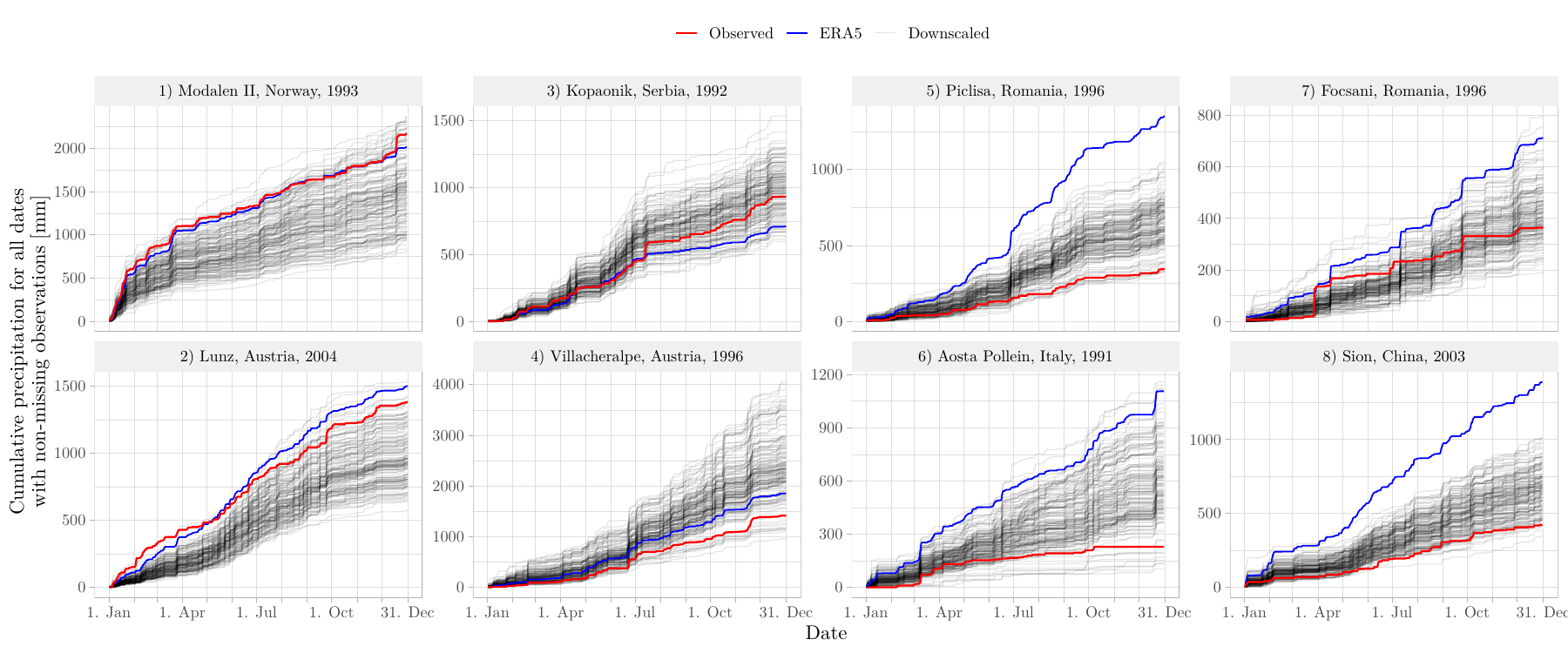}
  \caption{%
    Yearly cumulative precipitation for eight different combinations of weather stations and years, created using station observations (red line), ERA5 reanalysis data (blue line) and 150 simulated ensemble members from the full downscaling model (black lines).
    Some of the weather stations contains missing observations, so all cumulative precipitation values are computed using only dates with non-missing precipitation observations.
  }%
  \label{fig:precip_time_series}
\end{figure}

To further examine the properties of the full downscaling method, we plot time-series of simulated temperature and precipitation data for a selection of different years and weather stations, and compare them to observed data and ERA5.
Figure~\ref{fig:precip_time_series} displays cumulative precipitation data for eight different combinations of weather stations and years.
The four leftmost subplots represent weather stations where our downscaling method is outperformed by ERA5 for almost all of the chosen scores, while the four rightmost subplots represent the opposite.
In the four leftmost subplots, the general trend seems to be that the downscaled precipitation amounts are negatively or positively biased, while ERA5 is quite close to the truth.
However, even though the simulated precipitation amounts are biased, the observed precipitation amounts are still largely inside the range of the simulated ensemble members.
This implies that the simulated ensemble is well calibrated, even though it is biased.
For the four rightmost subplots of Figure~\ref{fig:precip_time_series}, the general trend seems to be that ERA5 overestimates the total precipitation amount, and that the simulated ensemble corrects for this bias.

\begin{figure}[t]
  \centering
  \includegraphics[width=.95\linewidth]{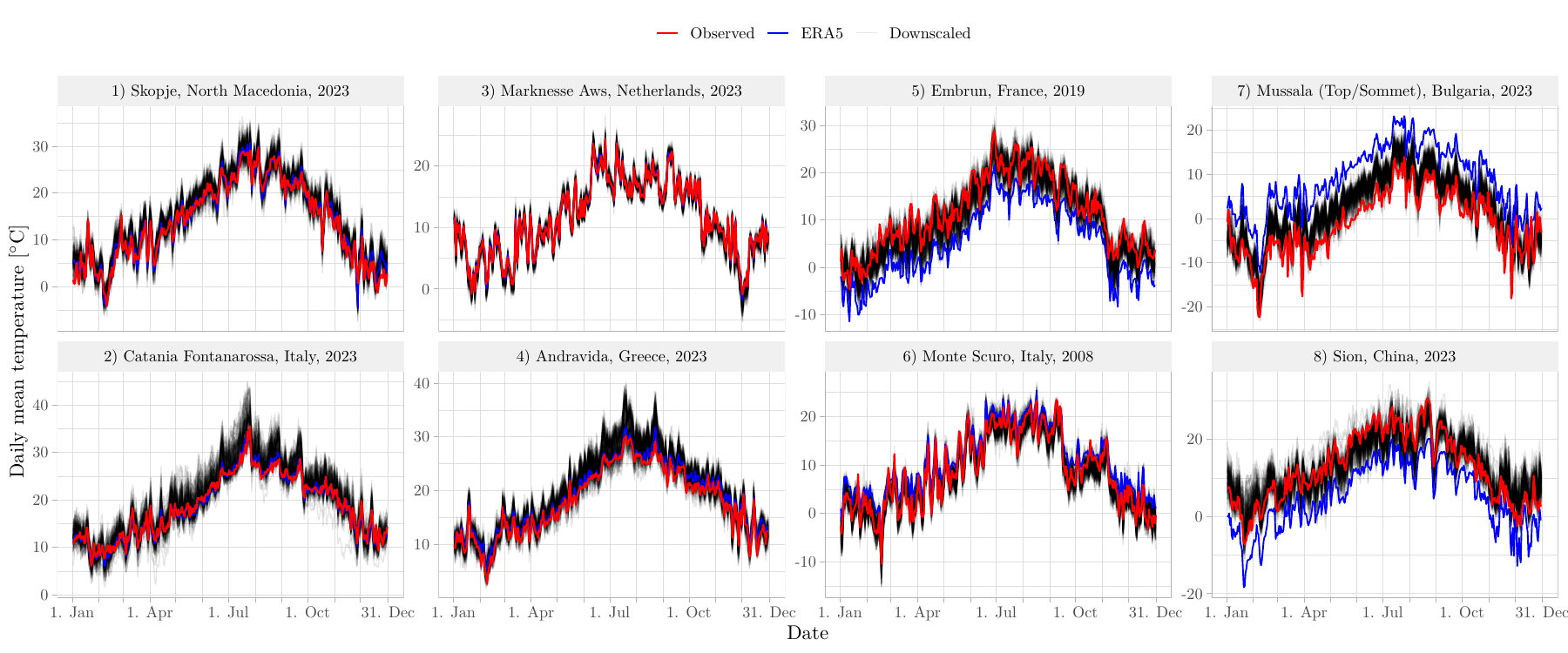}
  \caption{%
    Daily mean temperature for eight different combinations of weather stations and years, created using station observations (red line), ERA5 reanalysis data (blue line) and 150 simulated ensemble members from the full downscaling model (black lines).
  }%
  \label{fig:temp_time_series}
\end{figure}

Similarly to Figure~\ref{fig:precip_time_series}, Figure~\ref{fig:temp_time_series} displays examples of simulated temperature for four stations where our downscaling method is outperformed by ERA5 (left side), and four stations where our downscaling method outperforms ERA5 (right side).
Here, the differences between observed data and reanalysis data are considerably smaller, as temperature is a smoother process than precipitation.
The overall trend for the four leftmost subplots seems to be that the downscaled ensemble is slightly biased, with a possibly too large spread.
For the four rightmost subplots, there is a clear bias in the reanalysis temperatures, which is properly corrected by the downscaling.

\subsection{CPRCM comparison}%
\label{sec:cprcm}

\begin{figure}[t]
  \centering
  \includegraphics[width=.95\linewidth]{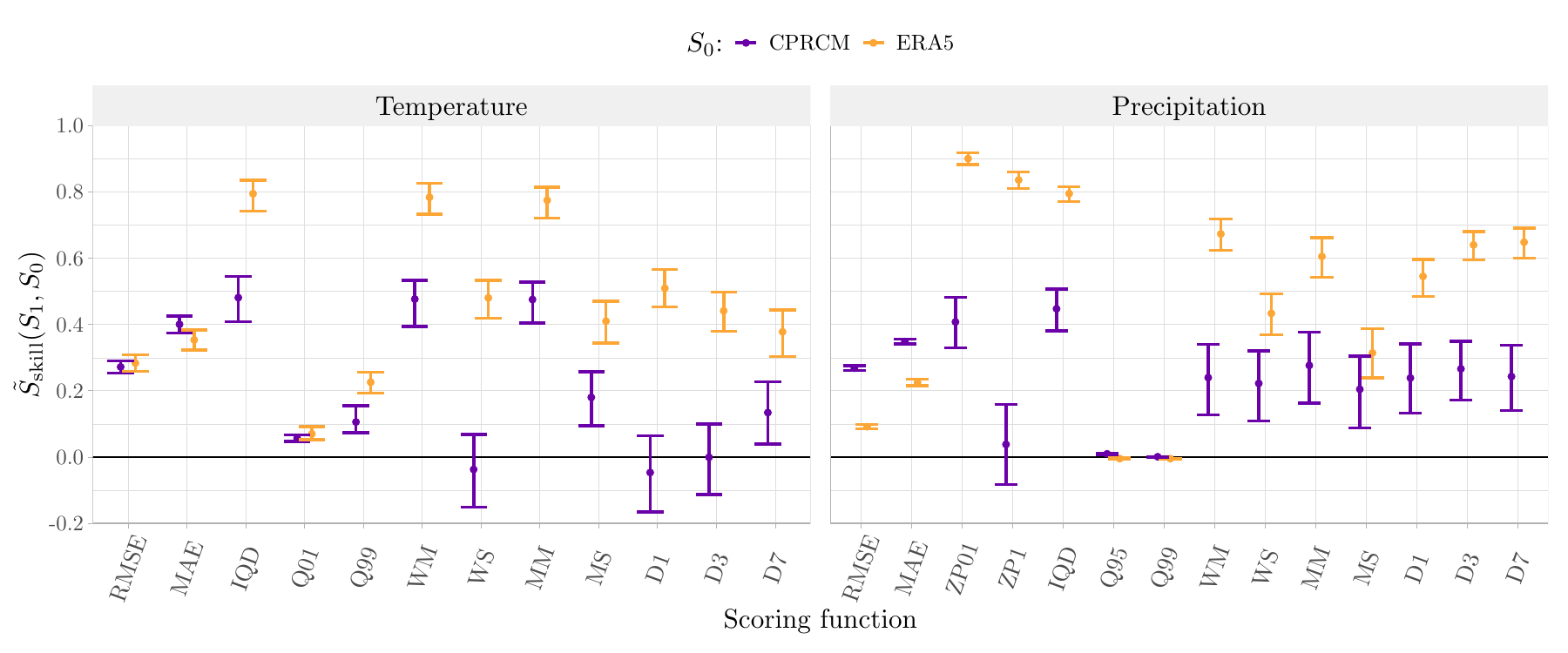}
  \caption{Skill scores comparing the full downscaling models against the CPRCM and ERA5, with bootstrapped \(95\%\) confidence intervals, for all evaluation criteria from Table~\ref{tab:evaluation}.
    The skill scores are computed with the full downscaling models as the competing models (\(S_1\)) and ERA5 or the CPRCM as base models (\(S_0\)).
    The left subplot displays results for the temperature models while the right subplot displays results for the precipitation models.%
}%
  \label{fig:cprcm_scores}
\end{figure}

Overall skill scores, comparing our full downscaling methods against both the CPRCM and ERA5, inside the ALP-3 domain, are displayed in Figure~\ref{fig:cprcm_scores}.
The CPRCM appears to outperform ERA5 for most evaluation criteria, except RMSE, MAE and some of the quantile scores.
However, the full downscaling models also outperform the CPRCM for most of the chosen criteria.

Even though the full downscaling models overall outperform the CPRCM, the results are not consistent across all locations and areas. 
This is further examined in Appendix~\ref{app:evaluation}.
We are unable to find any clear patterns, neither in space nor with respect to a range of different covariates, that describe the overall changes in skill between different weather stations.

\section{Discussion}%
\label{sec:discussion}

In this work, we downscaled precipitation and temperature from the ERA5 data product, which has a spatial resolution of \(0.25^\circ \times 0.25^\circ\).
Another alternative would have been to downscale the same variables from the ERA5-Land data product~\citep{munoz19_era5_land}, which has an even higher resolution of \(0.1^\circ \times 0.1^\circ\).
However, ERA5-Land only covers grid cells with a high enough land cover fraction, which means that there exist weather stations for which it would not be possible to directly apply our downscaling method.
For this reason, we chose to use the ERA5 data product, which covers the entire globe.
It should be trivial to apply our method for downscaling from ERA5-Land or other similar gridded historical weather data products.

The results in Sect.~\ref{sec:results} show that our stochastic downscaling model substantially outperforms a state-of-the-art dynamical downscaling model, i.e., the CPRCM, at reproducing local daily temperature and precipitation data.
This is an impressive result, as the CPRCM is a highly complex model that requires considerable computational resources, both to be run and to be extended to other spatial domains.
Our downscaling model, on the other hand, is simple and relatively interpretable, fast and computationally efficient, and easy to apply for any spatial location on Earth.
It should be noted that many other CPRCM simulations exists, also for other spatial and temporal domains.
It might be that other such downscaling methods perform better against our stochastic downscaling method with respect to the chosen evaluation criteria.
However, comparing our method to a large collection of different CPRCMs is outside the scope of this paper.

The evaluation approach in Sect.~\ref{sec:evaluation} is similar to leave-one-out cross-validation, as, for each \(i = 1, 2, \ldots, N\), no information from the \(i\)th local models are used for simulating data at the \(i\)th location.
However, it would be too computationally demanding to fit the global models to data from all but the \(i\)th location for each \(i = 1, 2, \ldots, N\).
We therefore allow a slight data contamination, by using the same global models for simulating data at each location.
The global models depend on data from thousands of different weather stations, while only relying on less than \(10\) covariates.
It therefore seems reasonable to assume that a global model fitted to data from all \(N\) locations is approximately identical to a global model fitted to data from \(N - 1\) locations.
For most practical purposes, our evaluation approach can therefore be considered as leave-one-out cross-validation.
We tested this assumption by fitting our global temperature model to data from \(N - 1\) locations, for \(10\) randomly selected locations.
The differences between the linear predictors of these \(10\) models, and between the linear predictor of the model fitted to all \(N\) locations, had an order of magnitude of \(0.01^\circ\)C.
This is so small that we do not believe it would have affected our model evaluation in a considerable way.

For, e.g., hydrological applications, the dependencies between local daily precipitation and temperature may be of importance.
Our method is unable to fully capture these, as precipitation and temperature are downscaled separately.
Parts of the dependence structure are still captured by our method, as the daily ERA5 temperature and precipitation values are used as covariates in all of the global and the local GAMs, for both variables.
However, to fully capture the internal variability between temperature and precipitation, the local time-series models would need to model the joint distributions of temperature and precipitation data.
Similarly, our method performs separate downscaling for each location of interest, and is unable to capture the internal variability between local weather at two nearby locations.
This is examined more closely in Appendix~\ref{app:spatial-coherence}.
There, we show that the downscaling models are able to capture some spatial properties of local data well, while failing to properly capture other spatial properties.
Further work on extending the time-series component of our model, using, e.g., multivariate ARMA models, state-space models or latent Gaussian models, might therefore make it possible to perform fully multivariate downscaling of both temperature and precipitation data, at multiple locations, jointly.
It might also be possible to turn our downscaled ensembles of temperature and precipitation into multivariate ensembles, using a high-dimensional version of the Schaake Shuffle \citep{ClarkEtAl2004SchaakeShuffleMethod}.
However, the Schaake Shuffle has mainly been used to reorder ensembles of point predictions, and we are not aware of any versions of the method that can be used to reorder ensembles of time series.

Our donor selection method only relies on locating the \(K\) nearest neighbouring stations.
This appears to work well overall, but it can result in some issues that might be fixed using a more complex donor selection method.
For stations located close to, or between, different climatological zones, the distance-based method may provide donor stations from highly different climatological zones, which may reduce the performance of the downscaling method.
As an example, weather stations along the Norwegian coastline may end up with both donor stations that are buoys, far into the North Sea, and high-altitude stations from mountainous areas.
This seems less optimal than simply selecting other coastal stations as donors.
A more sophisticated donor selection method for finding donors with more similar properties might therefore lead to improved downscaling in regions where small distances in space can lead to large difference in climate regimes.
One possible improvement to the donor selection method could be to transform all station locations into a climate space, following, e.g., \citet{CooleyEtAl2007BayesianSpatialModeling}, and to rely on distances in that space.
The number of donors might also be allowed to vary, based on climatological regimes, as more stations might be necessary to represent the overall variability of weather in certain regimes.
More complex, nonparametric and possibly deep-learning based transformations of the station locations might also be applied \citep[e.g.,][]{SampsonEtAl1992NonparametricEstimationNonstationary, ZammitMangionEtAl2021DeepCompositionalSpatial}, although these would considerably increase the complexity of the, otherwise relatively simple, downscaling method.

The addition of a first order Markov chain for modelling precipitation occurrences in the local models resulted in a considerable performance increase.
Based on this we also investigated the effects of exchanging the global occurrence model with a global first order Markov chain occurrence model.
This resulted in even further improvement with respect to the full precipitation downscaling model.
However, to keep the paper as concise as possible, and since the full precipitation downscaling model fits better into the overall model framework of the paper, and outperforms both ERA5 and the CPRCM, we chose to not focus on this extended model here.

\section{Conclusions}\label{sec:conclusions}

This paper proposes new models for downscaling time-series of daily temperature mean and precipitation from a gridded data set to any given location, based on a hierarchy of spline-based generalised additive models (GAMs) and auto-regressive moving average (ARMA) models.
For modelling precipitation occurrences, the GAMs are further organised in a way that results in a first order Markov chain structure.
The modelling framework utilises local information from donor locations in a neighbourhood around the location of interest, in addition to regional information.
The downscaling models are evaluated in a leave-one-out cross-validation study over a set of 4480 unique weather stations in Europe, and are shown to perform well compared to both ERA5 and a state-of-the-art convection-permitting dynamical downscaling method, with respect to a large collection of different evaluation criteria.
The downscaling models are also shown to outperform simpler downscaling models based on a single GAM, which are common choices for downscaling to locations without available high-resolution data.
Our results further stress the importance of evaluating downscaling methods on a variety of evaluation criteria, including criteria that specifically target higher order structures such as temporal autocorrelation.
Criteria such as RMSE and MAE, that focus on point-by-point comparisons of deterministic time series, are not able to detect model deficiencies in higher order model structures.  

\section*{Acknowledgements}

\subsection*{Funding}

  All authors gracefully acknowledge the support of Horizon Europe projects Impetus4Change (I4C, grant id.\ 101081555) and FAME (grant id.\ 101092639),
  the support of NordForsk through grant 138366 (Future Food Security in the Nordic-Baltic Region)
  and the support of the Research Council of Norway through grant 309562 (Climate Futures).

\subsection*{Conflict of interest}

The authors declare that they have no conflict of interest.

\subsection*{Code and data availability}

  The code used in this paper is freely available online at \url{https://github.com/NorskRegnesentral/downscaleToPoint}.
  The data used in this paper is freely available online at \url{https://cds.climate.copernicus.eu/}
  (ERA5 data), \url{https://www.ncei.noaa.gov/products/etopo-global-relief-model} (DEM data),
  \url{https://www.ncei.noaa.gov/data/global-summary-of-the-day/access/} (GSOD data) and from one of the many available ESGF nodes, such as \url{https://esgf-node.ipsl.upmc.fr/projects/esgf-ipsl/} (CPRCM data).
  All the necessary data has also been compiled into a common data set, freely available at \url{https://zenodo.org/records/19723606}~\citep{vandeskog_2026_data}.

\printbibliography

\appendix

\section{Evaluating spatial coherence}%
\label{app:spatial-coherence}

The developed downscaling models are built to simulate an ensemble of time series of temperature or precipitation at any given location.
They are not built to capture the joint dependence structure between local temperature and precipitation, or between local weather at multiple different locations.
Yet, spatial coherence of downscaled weather can, e.g., be important for many types of hydrological applications.
It is possible to create spatiotemporal ensembles of temperature and precipitation with the developed downscaling models, even though this is not the main purpose of the models.

To create a spatiotemporal ensemble of precipitation or temperature, we first apply our method to simulate \(B\) ensemble members, independently, at each of \(M\) different locations.
Then, we choose some random ordering of the \(B\) members at each location.
Finally, we match together ensemble member \(i\) from each location to form a spatiotemporal ensemble member, for \(i = 1, 2, \ldots, B\).
The skill of these ensembles relies heavily on the dominating type of variability of the local weather.
If the local weather is dominated by external variability, which can be captured by ERA5, then our models might be able accurately represent the spatial dependence of observed weather.
On the other hand, if local weather is dominated by internal variability that cannot be captured by ERA5, the simulated ensembles might perform poorly.

To examine the spatial coherence of our downscaled weather, we create spatiotemporal ensembles as explained above, and compare spatial statistics of the simulated data with spatial statistics of the observed data.
For each weather station, we create a circle with radius 100~km, centered around that station.
We then locate all other weather stations inside that circle.
For all dates where the circle contains observations from 8 or more stations, we compute five daily statistics: mean, median, minimum, maximum and standard deviation, of all temperature and precipitation observations inside the circle.
This results in daily time series of five different spatial statistics, for spatial domains centered around each of the \(N\) available weather stations.
We then compute similar statistics based on simulated spatiotemporal ensembles from the full downscaling models.

\begin{figure}[t]
  \centering
  \includegraphics[width=.99\linewidth]{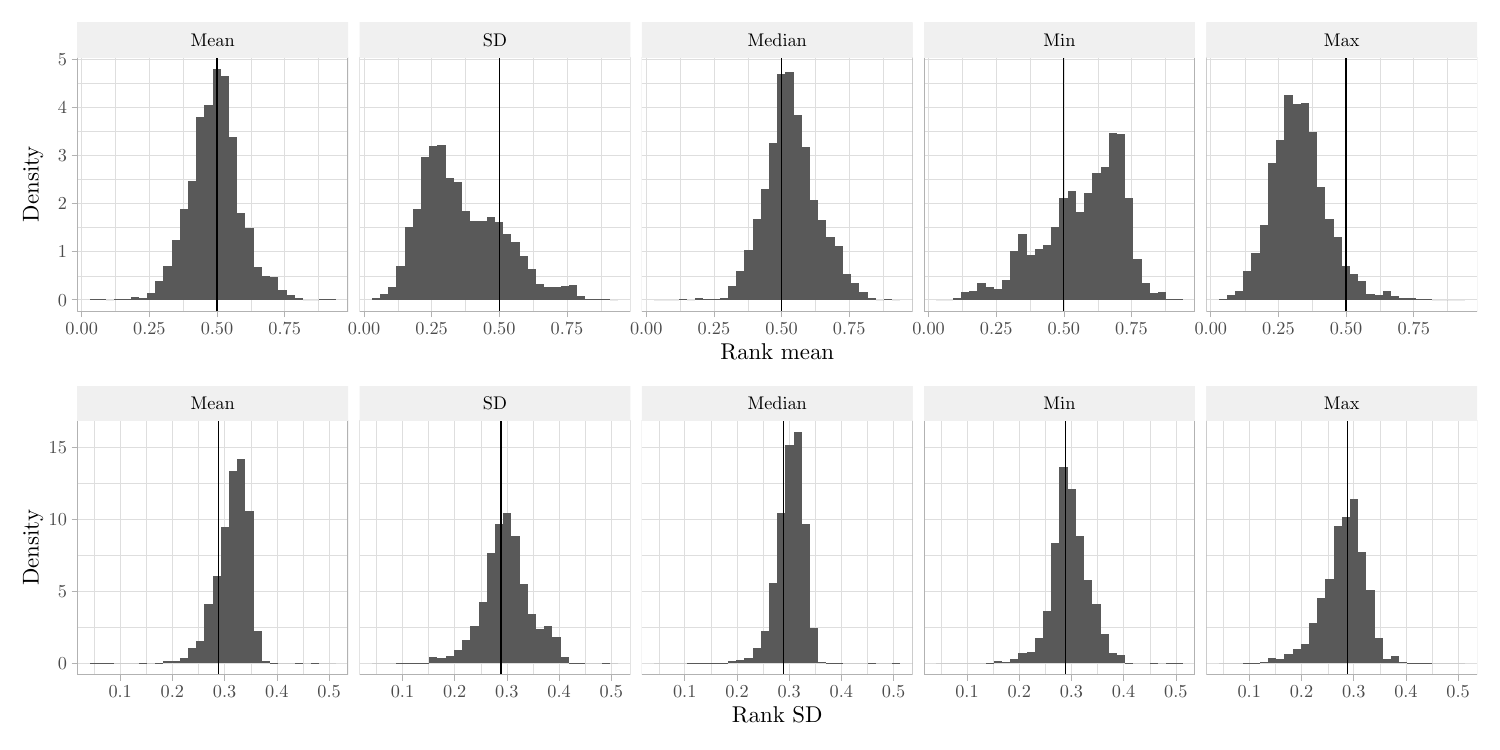}
  \caption{%
    Histograms displaying the overall rank means and rank standard deviations for all rank histograms for each of the five spatial temperature statistics.
}%
  \label{fig:temp_spatial_consistency}
\end{figure}

These simulated spatial statistics are evaluated in two ways.
First, we examine the calibration of the spatial statistics, by examining rank histograms \citep[e.g.,][]{ThorarinsdottirEtAl2016AssessingCalibrationHigh, VandeskogEtAl2022Quantilebasedmodeling}.
Second, we compute similar spatial statistics, based on ERA5 and the dynamically downscaled CPRCM data, and we compare the time series of the five spatial statistics in the same way that we compared time series of temperature and precipitation in Sect.~\ref{sec:evaluation}, using a subset of the evaluation criteria in Table~\ref{tab:evaluation}.

For each weather station, we create five rank histograms by finding the rank of each of the observed spatial statistics inside the ensemble of the \(B\) simulated spatial statistics.
If the downscaling model is perfectly calibrated, the rank histogram for each statistic and each station should be perfectly uniform.
In total, we create \(5 \times N\) rank histograms for each weather variable.
To aggregate the information from all these rank histograms, we standardise all the ranks to be a number between 0 and 1.
Then, we compute the mean and the standard deviation of each ranking histogram.
The uniform distribution between 0 and 1 has mean \(0.5\) and standard deviation \(1/\sqrt{12} \approx 0.29\).
We then examine how far away the rank histogram means and standard deviations are from \(0.5\) and \(1/\sqrt{12}\) to get an impression of the overall calibration of the simulated spatial statistics.
Figures~\ref{fig:temp_spatial_consistency} and~\ref{fig:temp_spatial_consistency} display histograms of the rank means and rank standard deviations for each of the five spatial statistics.
Overall, most of the spatial statistics seem to be fairly well calibrated, with most of their rank means clustering around \(0.5\) and most of their rank standard deviations clustering around \(1/\sqrt{12}\).
Other statistics attain poorer calibration, typically with too low rank means or too large rank standard deviations.
Too low rank means indicate that the simulations are biased and that they overestimate the spatial statistics.
Too large rank standard deviations indicate that the simulated spatial statistics are underdispersive, i.e., have too small variances.
The poorest calibration is found for some of the spatial precipitation statistics.
However, local precipitation is more difficult to model than local temperature, so it makes sense that spatially aggregated precipitation also is more difficult to model than spatially aggregated temperature.

\begin{figure}[t]
  \centering
  \includegraphics[width=.99\linewidth]{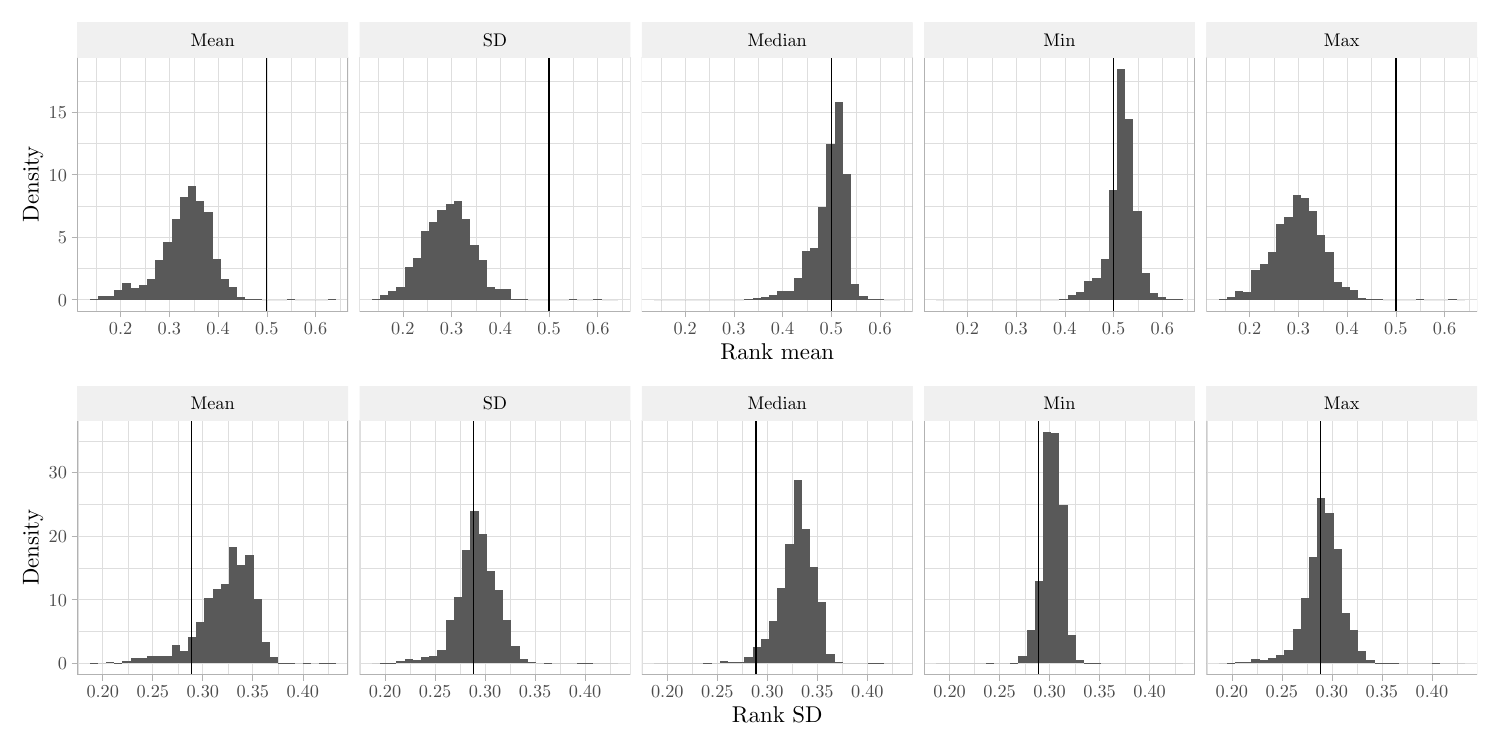}
  \caption{%
    Histograms displaying the overall rank means and rank standard deviations for all rank histograms for each of the five spatial precipitation statistics.
}%
  \label{fig:precip_spatial_consistency}
\end{figure}

\begin{figure}[tb]
  \centering
  \includegraphics[width=\linewidth]{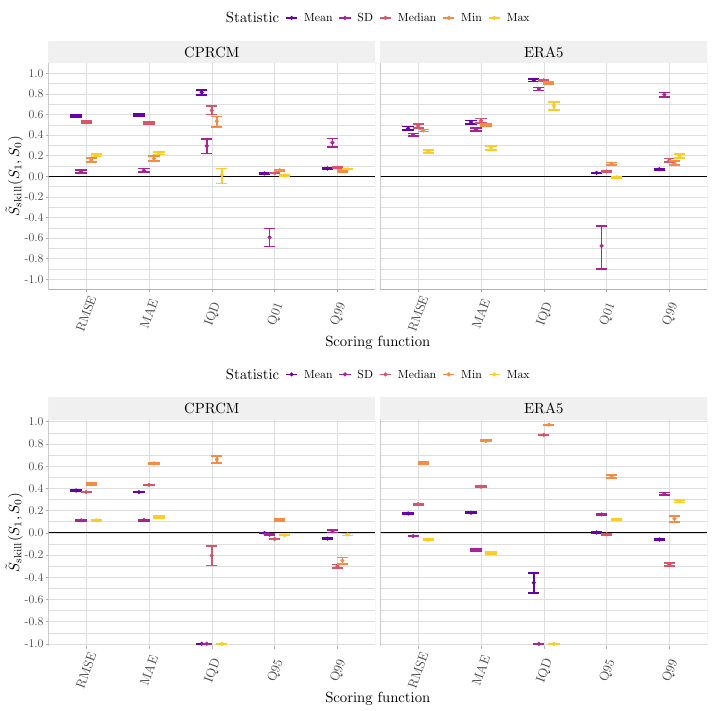}
  \caption{%
    Skill scores comparing the skill of the five different spatial statistics, for the full temperature (upper row) and precipitation (lower row) downscaling models, against ERA5 and the CPRCM.
    Error bars display bootstrapped \(95\%\) confidence intervals.
    The skill scores are created using a subset of the evaluation criteria in Table~\ref{tab:evaluation}, with the full downscaling model as the competitor (\(S_1\)).
    Skill scores for the five different spatial statistics are separated by different colours.
}%
  \label{fig:cprcm_spatial_scores}
\end{figure}

Figure~\ref{fig:cprcm_spatial_scores} displays overall skill scores for the five spatial statistics, created using all weather stations inside the ALP-3 domain.
The temperature downscaling model outperforms the CPRCM and ERA5 for most of the spatial statistics.
For the spatial precipitation statistics, the downscaling model often has higher skill than its competitors with respect to RMSE and MAE, but lower skill with respect to IQD and Q95/Q99.
This indicates that the full downscaling model is better at estimating the mean and median of the spatial statistics, but worse at estimating the overall distributions of the spatial statistics.
However, for most hydrological distributions, the main focus lies in estimating the actual spatial statistics, such as total rainfall over a catchment, and not the uncertainty distribution of that spatial statistic.
Thus, even though the full downscaling models are not developed with a focus on spatial modelling, they appear to outperform ERA5 and the CPRCM at estimating spatial statistics of local weather at multiple different locations.
We also compare the full downscaling models with ERA5 for the entire European region, visualised in Figure~\ref{fig:map}.
The results show similar patterns as displayed in Figure~\ref{fig:cprcm_spatial_scores}.

\section{Additional model evaluation}%
\label{app:evaluation}

\begin{figure}
  \centering
  \includegraphics[width=.95\linewidth]{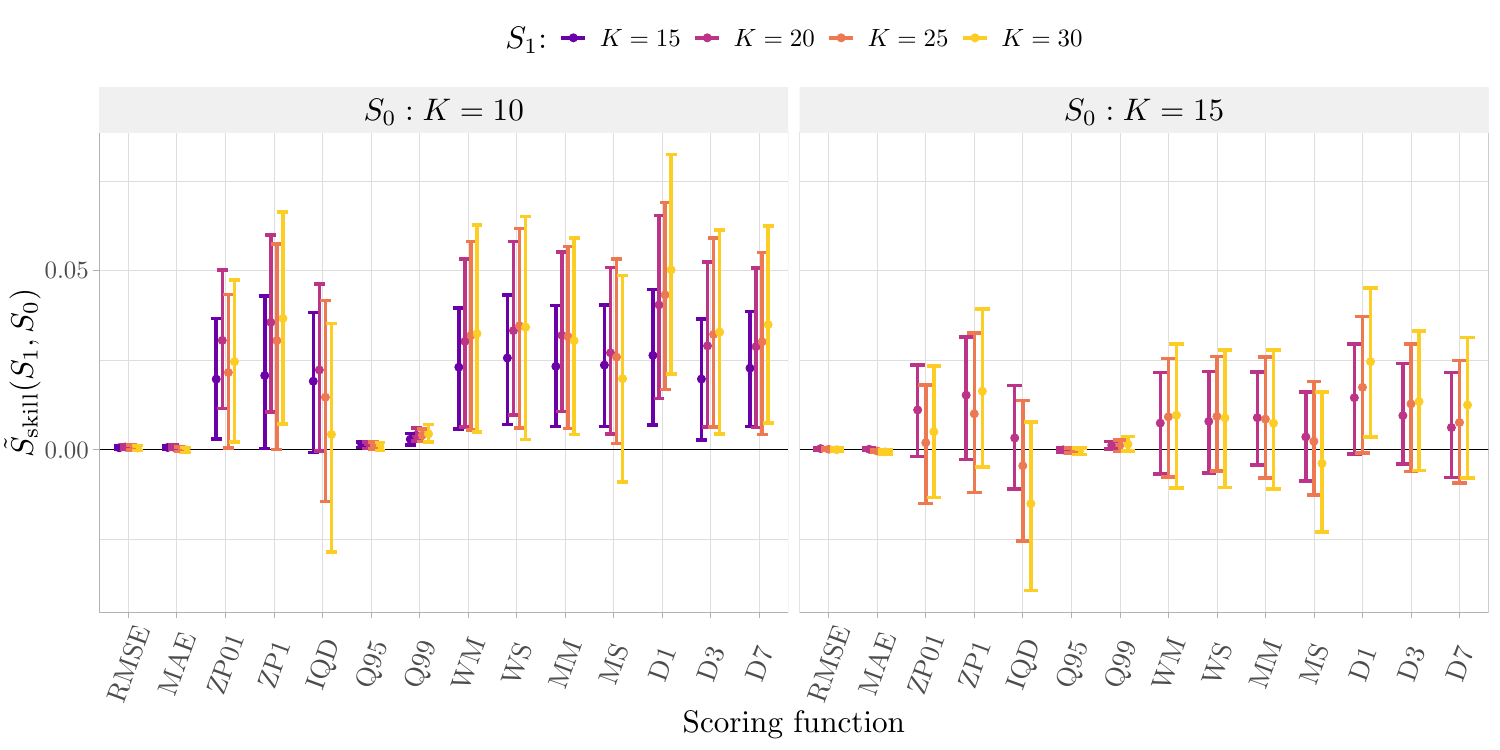}
  \caption{Skill scores with bootstrapped \(95\%\) confidence intervals, comparing different values of \(K\) for the full precipitation downscaling model.
    Competitor models (\(S_1\)) are distinguished by different colours, while base models (\(S_0\)) are distinguished by the different subplots.%
  }%
  \label{fig:precip_K_scores}
\end{figure}

\begin{figure}
  \centering
  \includegraphics[width=.95\linewidth]{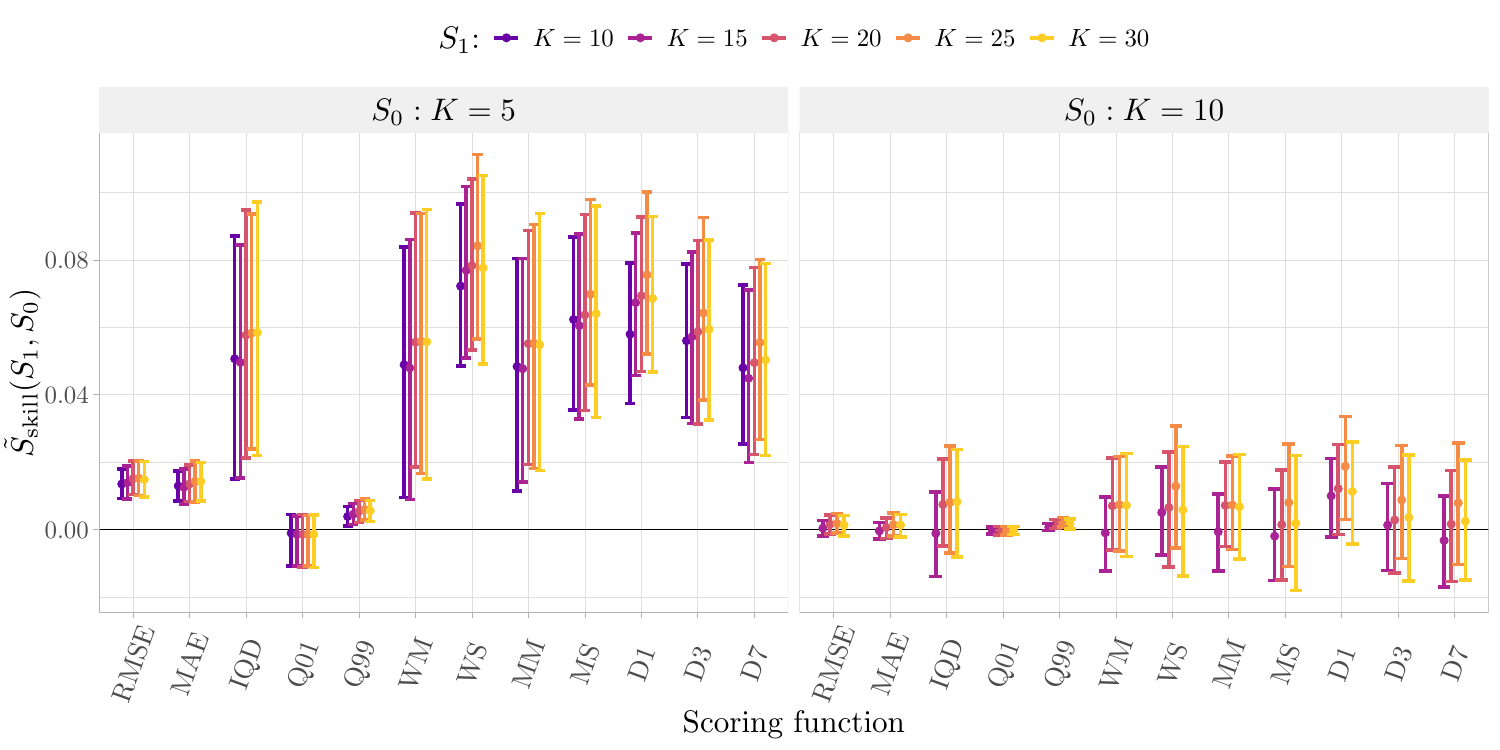}
  \caption{Skill scores with bootstrapped \(95\%\) confidence intervals, comparing different values of \(K\), for the full temperature downscaling model.
    Competitor models (\(S_1\)) are distinguished by different colours, while base models (\(S_0\)) are distinguished by the different subplots.%
  }%
  \label{fig:temp_K_scores}
\end{figure}

Figures~\ref{fig:precip_K_scores} and~\ref{fig:temp_K_scores} display all skill scores from Table~\ref{tab:evaluation} for the full temperature and precipitation downscaling models, respectively, with different values of \(K\). 

The skill scores are computed by considering a model with a low value of \(K\) as the base model, and then treating the same model with a higher value of \(K\) as a competing model.
Figure~\ref{fig:precip_K_scores} shows that the skill scores for describing precipitation occurrences and 1-day differences are positive when comparing \(K = 15\) with \(K = 10\).
However, after \(K = 15\), there does not seem to be any more skill to gain by further increasing the number of donors.
We therefore choose \(K = 15\) for downscaling precipitation.
The results for the local precipitation downscaling model are similar to that of the full model.

Figure~\ref{fig:temp_K_scores} shows that there is a considerable gain in skill when increasing the number of donors from \(K = 5\) to \(K = 10\) for the temperature downscaling model.
After that, however, there is little to be gained from further increasing \(K\).
We therefore choose \(K = 10\).
The results for the local deterministic temperature downscaling model is similar to that of the full model.

\begin{figure}[tb]
  \centering
  \includegraphics[width=.95\linewidth]{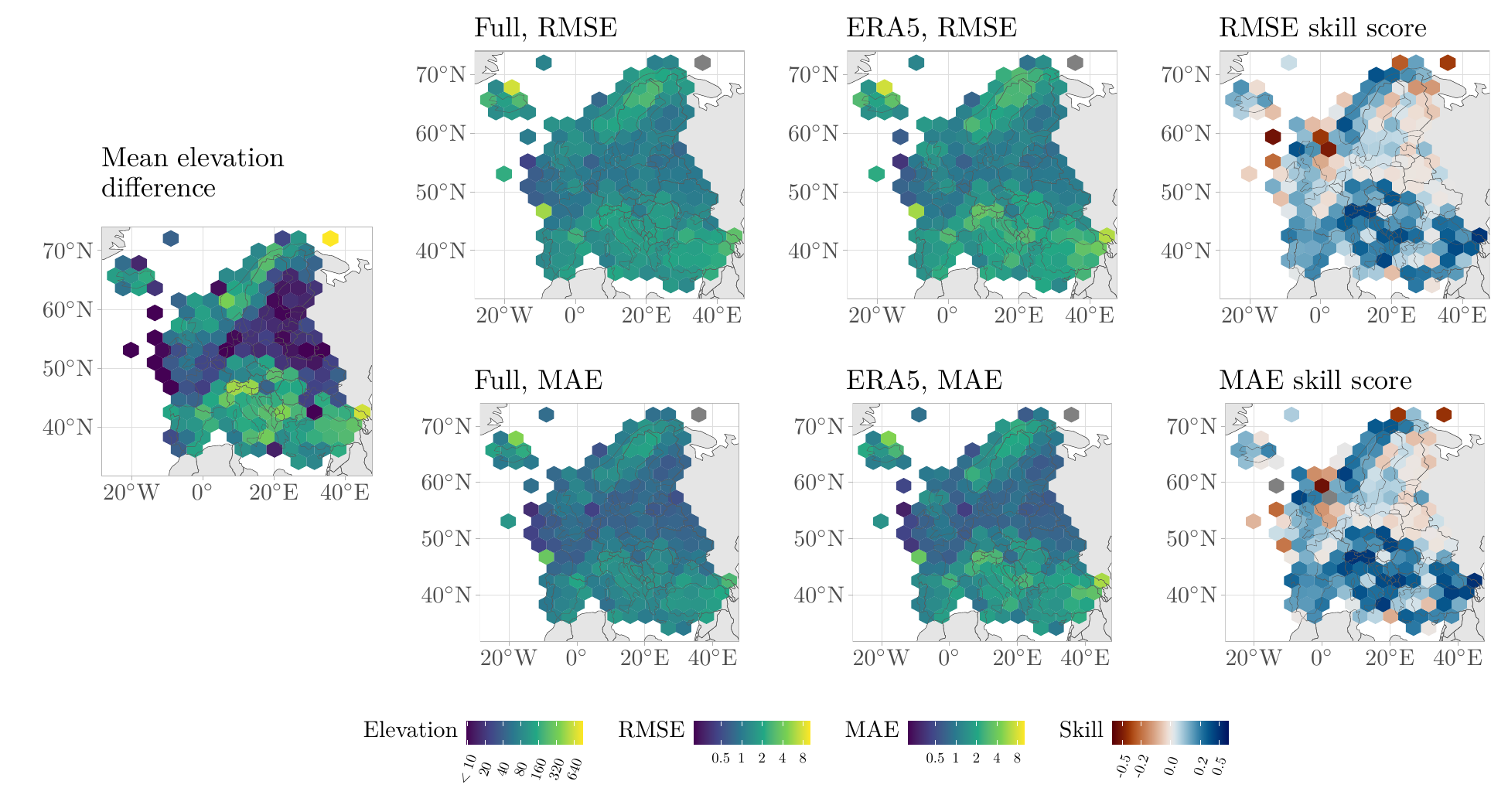}
  \caption{%
    Maps displaying the average temperature RMSE/MAE scores and skill scores for ERA5 and the full downscaling model, for all weather stations inside each hexagon of an hexagonal tiling of Europe.
    The skill scores are computed with the full precipitation downscaling model, against ERA5 as the base model.
    Additionally, the leftmost subplot displays the average mean elevation difference between ERA5 and the available weather stations within each hexagon.%
  }%
  \label{fig:temp_map_scores}
\end{figure}

Figure~\ref{fig:temp_map_scores} displays maps of average temperature RMSE and MAE scores for ERA5 and the full downscaling model, and the corresponding skill scores to compare these two methods, within a hexagonal tiling of space.
The figure also contains a map of average elevation differences between weather stations and ERA5 grid cells for the same hexagonal tiling.
There is a strong correlation between the average elevation differences and the average scores and skill scores within the hexagons.
The areas with large elevation differences are also the areas where our downscaling method achieves the highest skill scores with respect to ERA5.
Figure~\ref{fig:temp_map_scores} also displays that many of the areas where the full downscaling method performs worst compared to ERA5 are coastal or maritime areas.

\begin{figure}[tb]
  \centering
  \includegraphics[width=.95\linewidth]{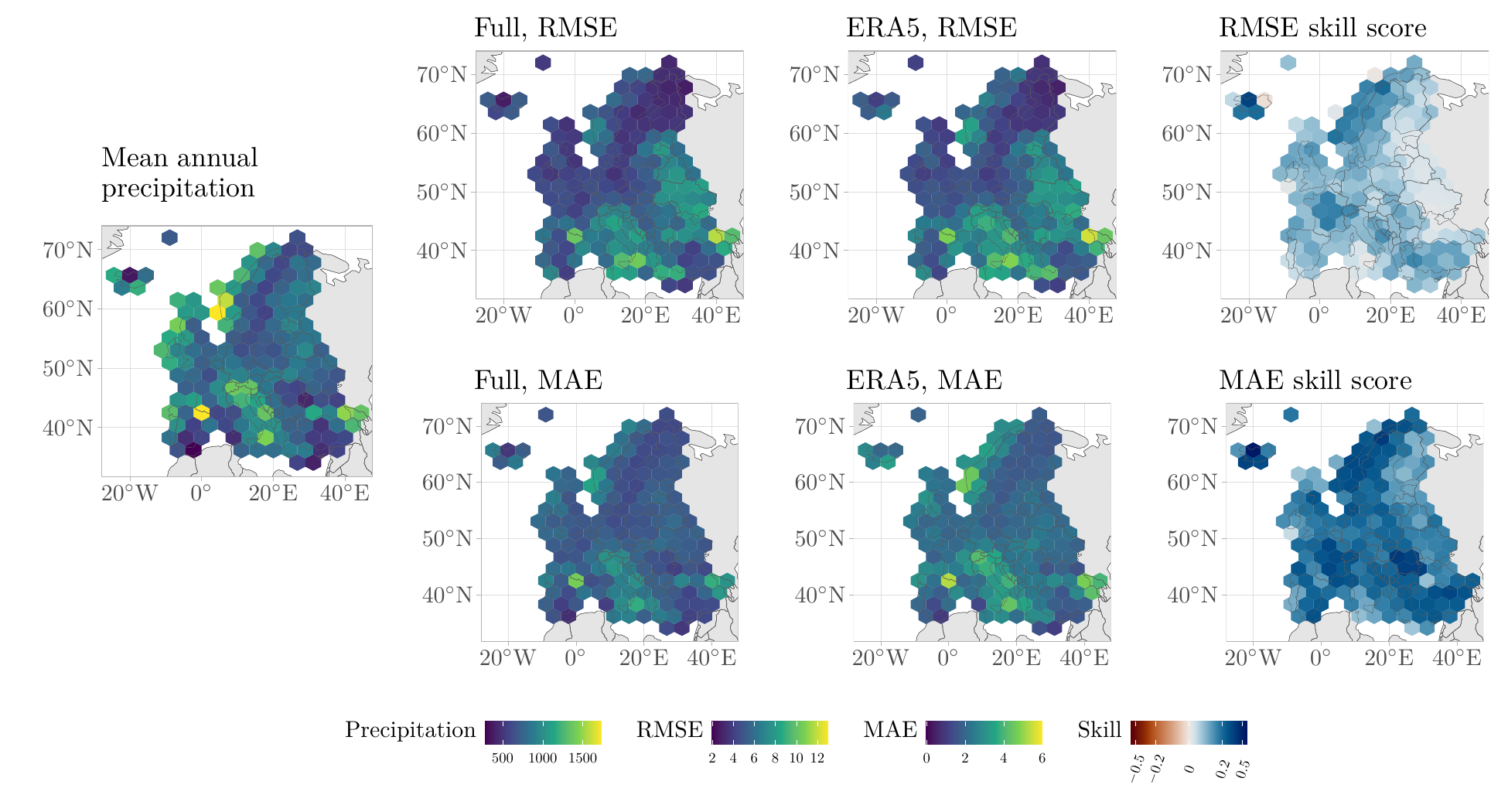}
  \caption{%
    Maps displaying the average precipitation RMSE/MAE scores and skill scores for ERA5 and the full downscaling model, for all weather stations inside each hexagon of an hexagonal tiling of the plane.
    The skill scores are computed with the full precipitation downscaling model, against ERA5 as the base model.
    Additionally, the leftmost subplot displays the average mean annual precipitation for all weather stations within each hexagon.%
  }%
  \label{fig:precip_map_scores}
\end{figure}

Figure~\ref{fig:precip_map_scores} displays similar maps as in Figure~\ref{fig:temp_map_scores}, but for precipitation instead of temperature.
The figure also contains a map of the mean annual precipitation for all weather stations within each spatial hexagon.
There is a strong correlation between mean annual precipitation and RMSE/MAE.
The skill score maps in Figure~\ref{fig:precip_map_scores} are more uniformly positive than those in Figure~\ref{fig:temp_map_scores}, and we do not find the same correlation between skill and distance to the ocean in this instance.

\begin{figure}[tb]
  \centering
  \includegraphics[width=.95\linewidth]{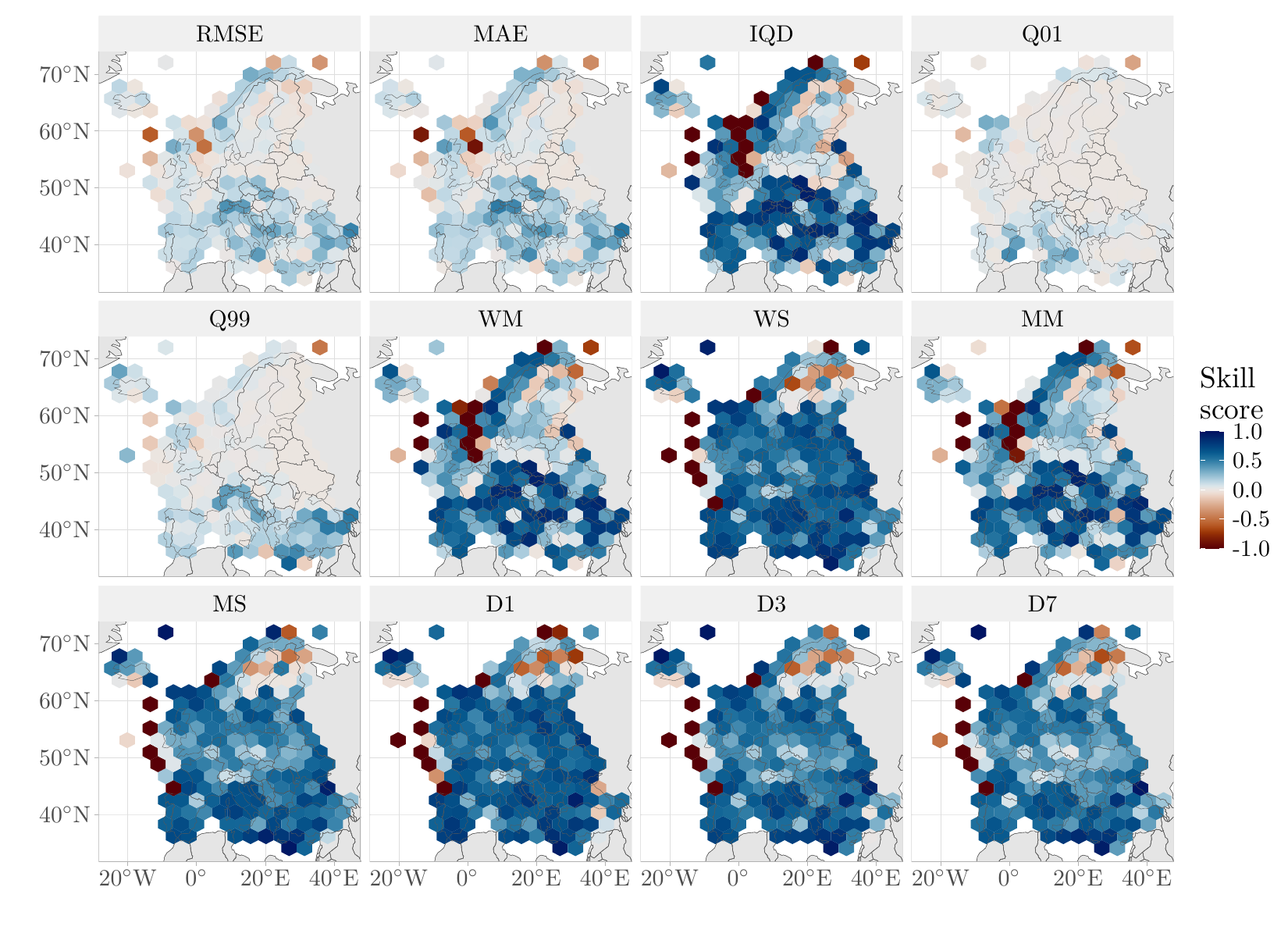}
  \caption{%
    Maps displaying the average temperature skill scores for all weather stations inside each hexagon of an hexagonal tiling of the plane.
    The skill scores are computed with the full temperature downscaling model with \(K = 10\) as the competitor, against ERA5 as the base model.
    Each subplot displays skill scores created using one of the scores described in Table~\ref{tab:evaluation}.%
  }%
  \label{fig:temp_map_scores_all}
\end{figure}

\begin{figure}[tb]
  \centering
  \includegraphics[width=.95\linewidth]{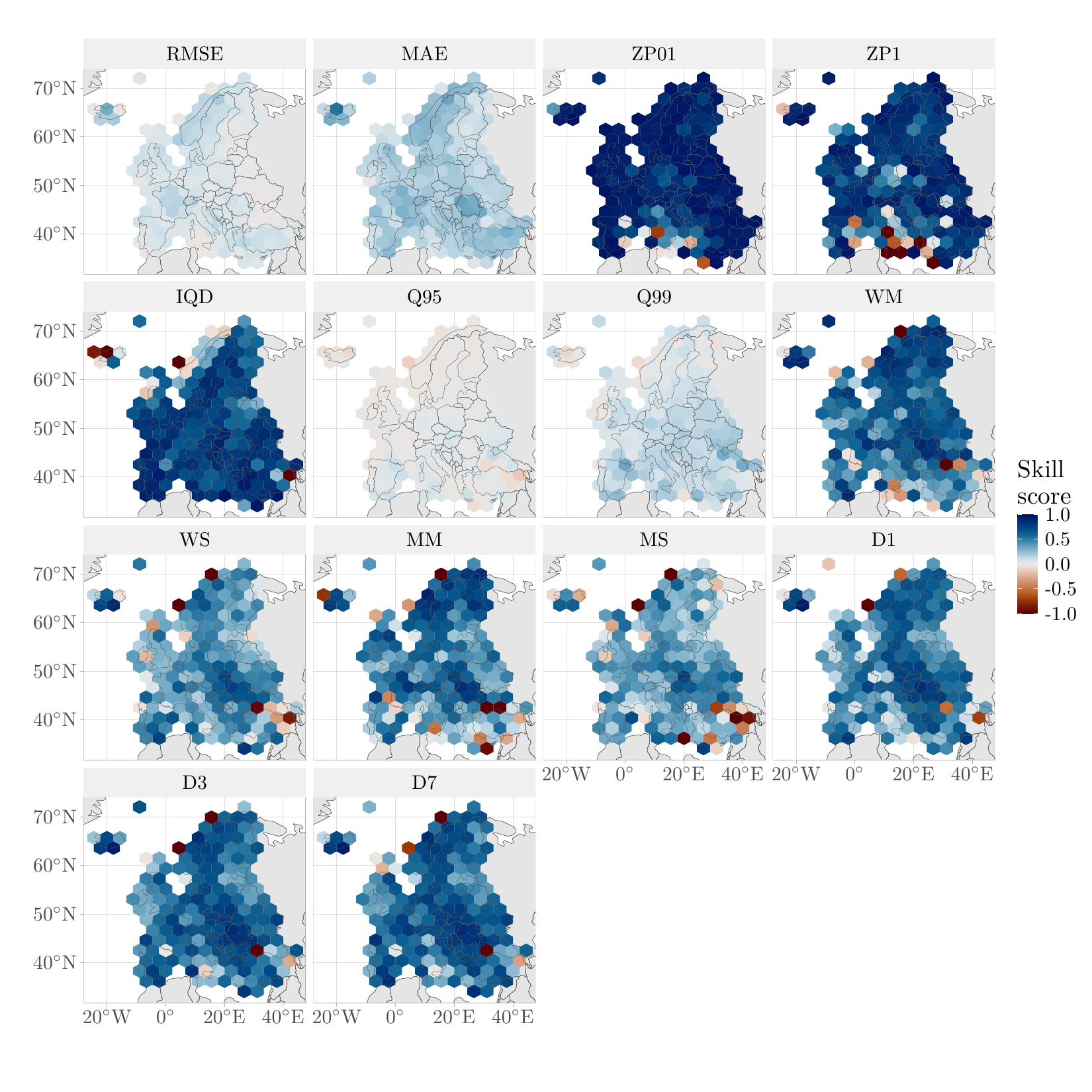}
  \caption{%
    Maps displaying the average precipitation skill scores for all weather stations inside each hexagon of an hexagonal tiling of the plane.
    The skill scores are computed with the full precipitation downscaling model with \(K = 15\) as the competitor, against ERA5 as the base model.
    Each subplot displays skill scores created using one of the scores described in Table~\ref{tab:evaluation}.%
  }%
  \label{fig:precip_map_scores_all}
\end{figure}

\begin{figure}[tb]
  \centering
  \includegraphics[width=.95\linewidth]{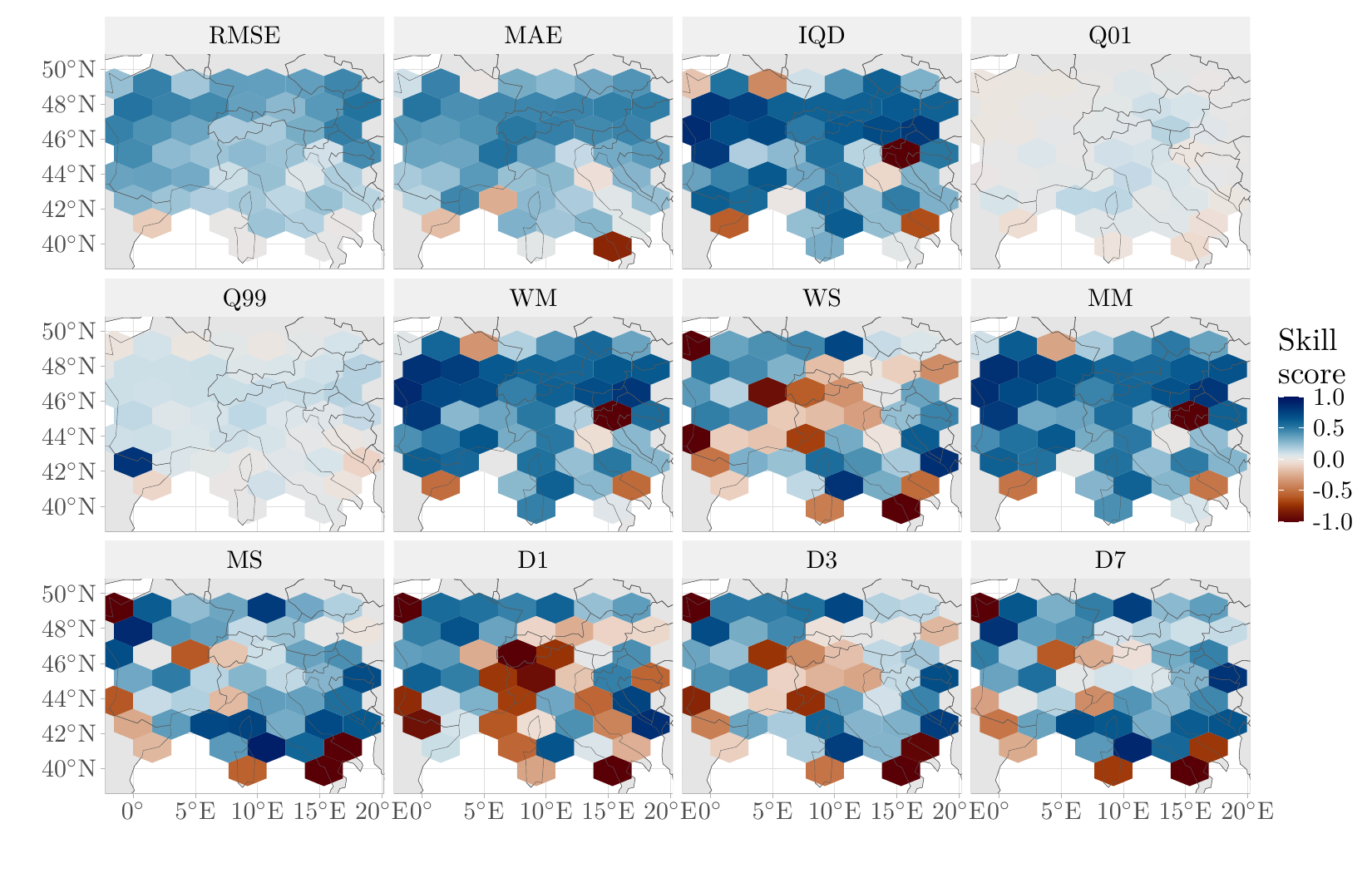}
  \caption{%
    Maps displaying the average temperature skill scores for all weather stations inside each hexagon of an hexagonal tiling of the plane.
    The skill scores are computed with the full precipitation downscaling model, with \(K = 10\) as the competitor, against the CPRCM as the base model.
    Each subplot displays skill scores created using one of the scores described in Table~\ref{tab:evaluation}.%
  }%
  \label{fig:temp_map_scores_cprcm_all}
\end{figure}

\begin{figure}[tb]
  \centering
  \includegraphics[width=.95\linewidth]{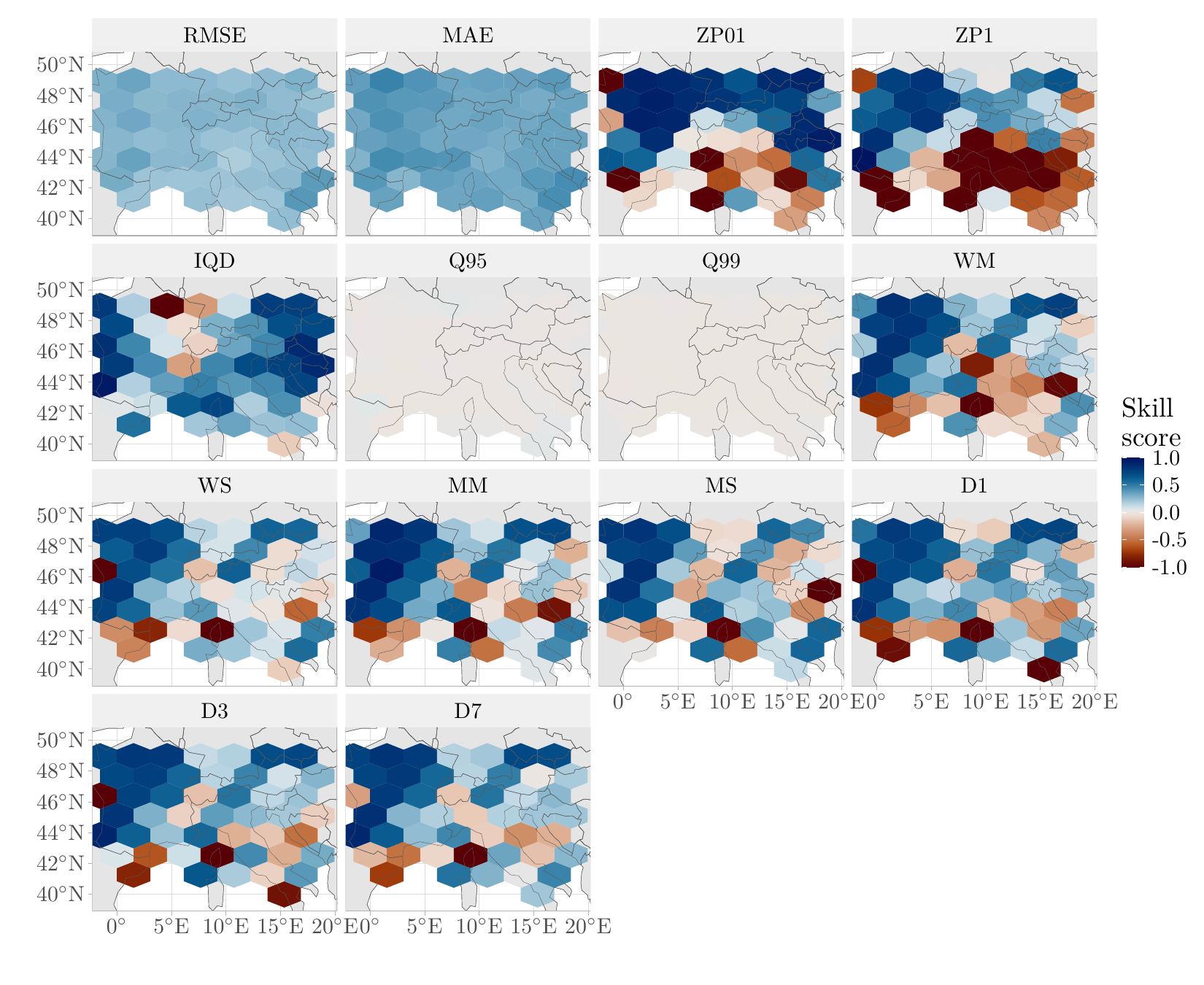}
  \caption{%
    Maps displaying the average precipitation skill scores for all weather stations inside each hexagon of an hexagonal tiling of the plane.
    The skill scores are computed with the full precipitation downscaling model, with \(K = 15\), as the competitor, against the CPRCM as the base model.
    Each subplot displays skill scores created using one of the scores described in Table~\ref{tab:evaluation}.%
  }%
  \label{fig:precip_map_scores_cprcm_all}
\end{figure}

Figures~\ref{fig:temp_map_scores_all} and~\ref{fig:precip_map_scores_all} display similar skill scores as in Figures~\ref{fig:temp_map_scores} and~\ref{fig:precip_map_scores}, for a much larger selection of evaluation criteria, with the full downscaling models as the competitor models and ERA5 as the base models.
Similarly, Figures~\ref{fig:temp_map_scores_cprcm_all} and~\ref{fig:precip_map_scores_cprcm_all} display skill score maps for the full downscaling models as the competitors and the CPRCM as the base model.

Figure~\ref{fig:temp_map_scores_all} show that, although the full temperature downscaling method generally outperforms ERA5, we sometimes experience poor skill in maritime areas.
There are multiple possible explanations for why this happens.
Coastal and maritime regions often tend to be flatter, with fewer elevation changes than inland regions.
As demonstrated in Sec.~\ref{sec:results}, large elevation differences within an ERA5 grid cell can lead to large biases in the ERA5 weather variables.
We therefore expect ERA5 to perform better in regions with small elevation differences, which, in turn, might reduce the skill of our downscaling method compared to ERA5.
Additionally, the ERA5 reanalysis is aware of the amount of land and ocean within each grid cell, and it accounts for the fact that land-atmosphere interactions can be considerably different than ocean-atmosphere interactions.
The connections between ERA5 and local weather might therefore differ between inland and coastal/maritime climates, and we might improve performance by incorporating additional covariates into our downscaling models, such as the distance to the sea, the percentage of land cover within each ERA5 grid cell, or a categorical variable that can distinguish between different types of climatic areas.
Finally, our framework relies on using information from the \(K\) nearest available weather stations when performing downscaling.
However, in coastal and maritime areas, the \(K\) nearest weather stations might include both buoys from far into the ocean and weather stations from areas further inland.
Since the connections between local weather and ERA5 might differ considerably between such different types of weather stations, it could, e.g., be problematic to use too many buoy donors for downscaling to a coastal or inland location, and vice versa.

Figures~\ref{fig:temp_map_scores_cprcm_all} and~\ref{fig:precip_map_scores_cprcm_all} provide a more nuanced understanding of the differences between the CPRCM and our full downscaling models.
They show that, although the full downscaling models overall outperform the CPRCM with respect to most of the chosen evaluation criteria, there are considerable spatial areas where the opposite holds.
We are unable to find any clear spatial patterns to describe the overall changes in skill scores for all evaluation criteria and for the two downscaling models.

\end{document}